\documentclass[journal]{IEEEtran}

\usepackage{ifpdf}

\usepackage{cite}

\ifCLASSINFOpdf
  \usepackage[pdftex]{graphicx}  
\else
  \usepackage[dvips]{graphicx}
\fi
\graphicspath{{ }}

\usepackage{amsmath}     
\usepackage{enumitem}    
\usepackage{adjustbox}   
\usepackage{hyperref}    
\usepackage{placeins}    
\usepackage{subcaption}  
\usepackage{float}       
\usepackage{caption}     
\usepackage{dblfloatfix} 
\usepackage{booktabs}
\usepackage{tikz}
\usetikzlibrary{arrows.meta,positioning,shapes.geometric,fit,backgrounds}
\usepackage{amssymb}
\usepackage[compatibility=false]{caption}
\usepackage{graphicx}
\usepackage{subcaption} 
\usepackage{booktabs}
\usepackage[space]{grffile}
\usepackage{multirow}
\usepackage{makecell}
\usepackage{array}
\usepackage[table,xcdraw]{xcolor}
\usepackage{booktabs}
\usepackage{siunitx} 

\usepackage{amsmath}

\usepackage{url}

\begin{document}
\hyphenation{op-tical net-works semi-conduc-tor}

\title{AmbiDrop: Ambisonics-Based Array-Agnostic Neural Speech Enhancement}

\author{Michael Tatarjitzky,~\IEEEmembership{Student~Member,~IEEE,} 
        Vladimir~Tourbabin,~\IEEEmembership{Member,~IEEE,} and~Boaz~Rafaely,~\IEEEmembership{Fellow,~IEEE}%
\thanks{Michael Tatarjitzky and Boaz Rafaely are with the School of Electrical and Computer Engineering, Ben-Gurion University of the Negev, Beer-Sheva, Israel.}
\thanks{Vladimir Tourbabin is with Reality Labs Research at Meta, Redmond, WA, USA.}%
}

%
%

%

\maketitle

\begin{abstract}
Multichannel Deep Neural Networks (DNNs) have significantly improved speech enhancement performance; however, they typically remain constrained by reliance on fixed microphone array geometries, leading to poor generalization on unseen or irregular configurations. Current array-agnostic approaches often rely on high-complexity architectures or massive, diverse datasets, yet they still struggle to generalize to out-of-distribution layouts. In this paper, we present an in-depth analysis of \textbf{AmbiDrop}, a recently proposed framework that achieves geometry independence by leveraging ideal Ambisonics as the DNN input. By employing a channel-wise dropout layer during training to simulate Ambisonics encoding errors, AmbiDrop decouples the learning process from the physical sensor arrangement. During inference, microphone signals from arbitrary array configurations are transformed into the Ambisonics domain via Ambisonics Signal Matching (ASM) before processing. Extensive experiments demonstrate that AmbiDrop maintains high robustness across a diverse suite of unseen simulated arrays and real-world recordings. Furthermore, our results show that the framework is resilient to sensor failures and remains effective even with reduced network scales, making it highly suitable for deployment on resource-constrained edge devices and versatile wearable hardware.
\end{abstract}


%
\IEEEpeerreviewmaketitle

\section{Introduction}
\label{sec:Introduction}
\IEEEPARstart{M}{odern} devices and wearables increasingly leverage multi-microphone arrays to perform sophisticated speech enhancement tasks, such as dereverberation, target speaker extraction, and noise reduction. While classical signal processing techniques—including beamforming \cite{VanTrees2002}, multichannel Wiener filtering \cite{modhave2016design}, and spectral masking \cite{minipriya2018review}—offer low algorithmic latency and predictable performance, they rely on simplified physical models that often degrade in complex, non-stationary acoustic environments.

To achieve greater robustness, Deep Neural Networks (DNNs) have become the preferred approach. By simultaneously exploiting temporal, spectral, and spatial cues, multichannel DNNs \cite{10819706, huang2025advances} significantly outperform classical methods in challenging acoustic scenes. However, a major limitation persists: these models are typically data-driven and optimized for a specific, fixed array geometry. Consequently, they often fail to generalize to real-world scenarios where recording devices vary in shape and microphone count, or when unexpected sensor failures occur.

To address these rigid constraints, array-agnostic DNNs have emerged as a promising research direction. These architectures aim to decouple the enhancement process from the specific sensor layout, providing a flexible solution for geometry-independent speech enhancement. Current approaches include the use of transform-average-concatenate (TAC) layers or their variations \cite{luo2020end, yoshioka2022vararray, jukic2023flexible, taherian2022one, zhang2021microphone}, which enable networks to remain invariant to the number and ordering of microphone channels. More complex architectures leverage attention-based modules \cite{lee2024deftan, wang2020neural}, such as spatial transformers or space-object cross-attention (SOCA) blocks, to better exploit spatial information for improved generalization. Furthermore, meta-learning techniques \cite{mannanova2024meta} have been proposed to increase robustness to array configuration changes, though these typically require highly diverse training sets. Additionally, recent eigenbeam-feature-based multi-order encoder \cite{11464064} attempts to achieve array-agnostic enhancement by transforming microphone inputs into a canonical circular harmonic space. Although this method demonstrates generalization across various planar configurations, it lacks evaluation on out-of-distribution (OOD) arrays that deviate from the training set.

Despite these advancements, significant challenges remain. Many existing methods either rely on the availability of massive, diverse array datasets for training or utilize complex architectures that still struggle with OOD geometries. Furthermore, the performance of these models is rarely validated across a comprehensive suite of arrays, ranging from familiar configurations to unusual or irregular geometries.

Recently, we proposed AmbiDrop \cite{tatarjitzky2026ambidrop}, a framework designed for array-agnostic speech enhancement, aiming to overcome the limitations of previous methods. AmbiDrop decouples the training process from specific array geometries by leveraging ideal Ambisonics signals \cite{zotter2019ambisonics} as input. These signals represent the sound field as a transformation of a plane-wave representation, which is independent of the recording device. Since each channel provides a spatially-consistent representation, Ambisonics signals are highly suitable for DNN training. However, obtaining perfect Ambisonics representations from physical arrays is challenging, particularly for non-spherical or sparse configurations \cite{rafaely2015fundamentals}. To account for the encoding errors with such arrays, AmbiDrop employs a channel-wise dropout layer during training to simulate imperfect spatial information. During inference, microphone signals are encoded into the Ambisonics domain using Ambisonics Signal Matching (ASM) \cite{gayer2024ambisonics} before being processed by the network.

In this work, we extend the preliminary feasibility study presented in \cite{tatarjitzky2026ambidrop} and present an in-depth investigation of the AmbiDrop framework, addressing several key gaps left open by the preceding conference paper:
\begin{itemize}
    \item We evaluate performance across a diverse set of simulated 1D, 2D, and 3D arrays using various Acoustic Transfer Function (ATF)  (\cite{tatarjitzky2026ambidrop} studied only 1D and 2D arrays).
    \item We examine the framework's generalizability by evaluating it with two distinct DNN architectures as backbones (\!\cite{tatarjitzky2026ambidrop} studied only one network backbone).
    \item We validate the system on real-world data using the Project Aria \cite{engel2023project} wearable glasses (\!\cite{tatarjitzky2026ambidrop} studied only simulated data).
    \item We conduct an extensive ablation study to analyze performance under different dropout configurations, missing input channels, and varying network scales (\!\cite{tatarjitzky2026ambidrop} did not study robustness to channel count, nor the issue of computation complexity).
\end{itemize}

The results of this paper demonstrate that by utilizing Ambisonics-aided array-free training, AmbiDrop maintains high robustness across diverse unseen array datasets and real-world recordings. In contrast, baseline geometry-agnostic DNNs trained with microphone signals as input and diverse array configurations, fail when faced with unseen configurations. Furthermore, we show that our framework remains resilient even when losing nearly half of the input microphones and maintains competitive performance with reduced network sizes. This versatility positions AmbiDrop as a robust framework for deployment across various DNN architectures and arbitrary microphone geometries.

\section{Background}
\label{sec:Background}
This section establishes the mathematical framework for the proposed method, detailing the signal model, the Ambisonics encoding process, and the core speech enhancement task formulation. We utilize a spherical coordinate system $(r, \theta, \phi)$ representing the radius, polar angle (inclination), and azimuth, respectively. Following the physics convention, the polar angle $\theta$ is measured from the positive $z$-axis (the north pole) downwards and ranges between $0$ and $\pi$. The wavenumber is defined as $k = \frac{2\pi}{c}f$, where $c$ denotes the speed of sound and $f$ represents the frequency.

\subsection{Array Signal Model}
\label{ssec:Array Signal Model}
Let an arbitrary array consist of $M$ omnidirectional microphones, where the position of each microphone is given by the coordinates $(r_m, \theta_m, \phi_m)$ for $m \in \{1, \dots, M\}$. We model the sound field as a superposition of $Q$ plane waves originating from a set of directions $\Omega_Q$, where each wave has its direction of arrival (DOA) $(\theta_q, \phi_q) \in \Omega_Q$. These plane waves represent the direct sound from sources and their respective reflections from the enclosure boundaries.

The array steering matrix is defined as $\mathbf{V}(k) \in \mathbb{C}^{M \times Q}$, where each entry $[\mathbf{V}(k)]_{m,q}$ is the frequency-dependent transfer function between the $q$-th plane wave and the $m$-th microphone. Consequently, the captured microphone signals in the frequency domain are modeled as:
\begin{equation}
    \mathbf{x}(k) = \mathbf{V}(k) \mathbf{s}(k) + \mathbf{n}(k),
    \label{eq:signal}
\end{equation}
where $\mathbf{x}(k) \in \mathbb{C}^{M \times 1}$ is the vector of microphone signals. The vector $\mathbf{s}(k) = [s_1(k), \dots, s_Q(k)]^T \in \mathbb{C}^{Q \times 1}$ contains the complex amplitudes of the $Q$ plane waves as measured at the coordinate origin. The additive noise component $\mathbf{n}(k)$ is assumed to be spatially white (i.i.d.) and uncorrelated with the source signals $\mathbf{s}(k)$.

\begin{figure*}[t]
\centering

\begin{subfigure}[t]{\linewidth}
  \centering
  \small
  \begin{tikzpicture}[
    node distance=8mm and 7mm,
    >=Latex,
    every node/.style={align=center},
    block/.style={draw, rounded corners=2pt, line width=0.4pt, fill=black!8,
                  minimum width=2.cm, minimum height=0.7cm, inner sep=2.5pt},
    arrow/.style={-{Latex[length=2.5mm,width=1.7mm]}, line width=0.5pt}
  ]
    \node (amb) {Ideal\\Ambisonics};
    \node[block, right=of amb] (drop) {Input Channel-wise\\Dropout};
    \node[block, right=of drop] (net) {DNN};
    \node[right=of net] (out) {Enhanced Speech};
    
    \draw[arrow] (amb) -- (drop);
    \draw[arrow] (drop) -- (net);
    \draw[arrow] (net) -- (out);
  \end{tikzpicture}
  \caption{Training phase}
  \label{fig:processing_chain_train}
\end{subfigure}

\vspace{4mm} 

\begin{subfigure}[t]{\linewidth}
  \centering
  \small
  \begin{tikzpicture}[
    node distance=8mm and 7mm,
    >=Latex,
    every node/.style={align=center},
    block/.style={draw, rounded corners=2pt, line width=0.4pt, fill=black!8,
                  minimum width=2.cm, minimum height=0.7cm, inner sep=2.5pt},
    arrow/.style={-{Latex[length=2.5mm,width=1.7mm]}, line width=0.5pt}
  ]
 \node (mic) {Microphone\\Signals};   
    \node [block, right=of mic](amb) {ASM};
    \node[block, right=of amb] (net) {DNN};
    \node[right=of net] (out) {Enhanced Speech};
    
    \draw[arrow] (mic) -- (amb);
    \draw[arrow] (amb) -- (net);
    \draw[arrow] (net) -- (out);
    
  \end{tikzpicture}
  \caption{Inference phase}
  \label{fig:processing_chain_inference}
\end{subfigure}

\caption{Schematic diagram of the proposed model: (a) training and (b) inference.}
\label{fig:processing_chain}
\end{figure*}
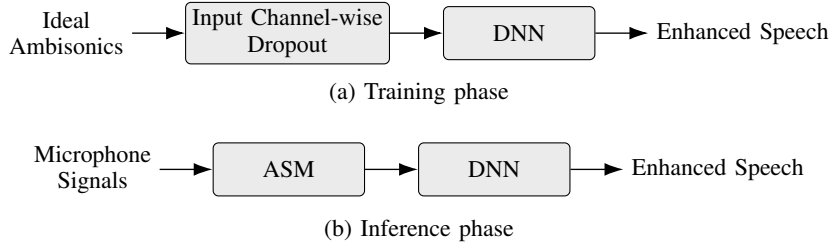

\subsection{Ambisonics Encoding}
\label{ssec:Ambisonics Encoding}
Ambisonics is a standard format \cite{zotter2019ambisonics} for representing a sound field using a Spherical Harmonic (SH) basis. This representation is based on the sound field excited by the source signals and is independent of the specific microphone array geometry. To define Ambisonics, we first introduce the complex SH basis functions:
\begin{equation}
    Y_n^m(\theta, \phi) \equiv \sqrt{\frac{2n+1}{4\pi} \frac{(n-m)!}{(n+m)!}} P_n^m(\cos \theta) e^{im\phi},
\end{equation}
where $P^m_n(\cdot)$ are the associated Legendre functions, $0 \leq n \leq N_a$ is the function order, and $-n \leq m \leq n$ is the function degree.

The Ambisonics signal vector is calculated as (see Eq. 2.43 in \cite{rafaely2015fundamentals}):
\begin{equation}
    \mathbf{a}_{nm}(k) = \mathbf{Y}^H_{\Omega_Q} \mathbf{s}(k).
\label{eq:amb}
\end{equation}
Here, $\mathbf{a}_{nm}(k)$ is the Ambisonics signal vector up to order $N_a$ with length $(N_a + 1)^2$, ordered according to the Ambisonics Channel Numbering (ACN) convention $n^2 + n + m$ with $0\le n \le N_a$ and $-n\le m \le n$. 
The matrix $\mathbf{Y}_{\Omega_Q} = [\mathbf{y}_{0,0}, \dots, \mathbf{y}_{N_a, N_a}] \in \mathbb{C}^{Q \times (N_a+1)^2}$ is the SH matrix, where each column vector $\mathbf{y}_{n,m} = [Y_n^m(\theta_1, \phi_1), \dots, Y_n^m(\theta_Q, \phi_Q)]^T$ of length $Q$ contains the SH basis functions evaluated at the $Q$ DOAs $(\theta_q, \phi_q)$ for $q=1, \dots, Q$.

As discussed later in \ref{sec:baseline}, certain Deep Neural Networks (DNNs) require real-valued input. By utilizing real-valued SH basis functions, which can be derived as shown in \cite{poletti2009unified}, real-valued Ambisonics signals can be computed using a formulation analogous to \eqref{eq:amb}. The resulting time-domain real-valued Ambisonics signals are denoted as $\mathbf{a}_{nm,\text{real}}(t)$.

Ambisonics Signal Matching (ASM) \cite{gayer2024ambisonics} provides a framework for encoding Ambisonics signals from arbitrary microphone array geometries. This is achieved through a linear transformation of the microphone observations into the $(n,m)$-th Ambisonics coefficient:
\begin{equation}
    \hat{a}_{nm}(k) = \mathbf{c}_{nm}^H \mathbf{x}(k),
    \label{eq:amb_est}
\end{equation}
where $0 \leq n \leq N_a$ and $-n \leq m \leq n$. The filter vector $\mathbf{c}_{nm}(k) \in \mathbb{C}^M$ is designed to minimize the normalized mean square error (NMSE) between the estimated and ground-truth Ambisonics coefficients, defined as:
\begin{equation}
\varepsilon_{\mathrm{Amb}}^{nm} (k) = 
\frac{E\left[ | \hat{a}_{nm}(k) - a_{nm}(k) |^2 \right]}
     {E\left[ | a_{nm}(k) |^2 \right]},
\label{eq:amb_error}
\end{equation}
where $E[\cdot]$ denotes the statistical expectation operator. By substituting \eqref{eq:signal}, \eqref{eq:amb}, and \eqref{eq:amb_est} into \eqref{eq:amb_error}, the NMSE can be expressed in terms of the array steering function. Under the assumption of a diffuse sound field—modeled by $Q$ uncorrelated plane waves with source covariance $\mathbf{R}_s(k) = \sigma_s^2 \mathbf{I}$—and spatially white noise with variance $\sigma_n^2$, the error simplifies to:
\begin{equation}
    \varepsilon_{\mathrm{Amb}}^{nm} (k) = \frac{\sigma_s^2 \left\| \mathbf{V}(k)^H \mathbf{c}_{nm}(k) - \mathbf{y}_{nm} \right\|_2^2 + \sigma_n^2 \left\| \mathbf{c}_{nm}(k) \right\|_2^2}{\sigma_s^2 \left\| \mathbf{y}_{nm} \right\|_2^2}.
    \label{eq:amb_error_simply}
\end{equation}
Minimizing \eqref{eq:amb_error_simply} with respect to $\mathbf{c}_{nm}(k)$ yields the optimal filter coefficients:
\begin{equation}
\mathbf{c}_{nm}^{\mathrm{opt}}(k) =
\left( \mathbf{V}(k)\mathbf{V}^H(k) + \frac{\sigma_n^2}{\sigma_s^2} \mathbf{I} \right)^{-1} 
\mathbf{V}(k) \mathbf{y}_{nm}.
\label{eq:c_opt}
\end{equation}
By substituting the optimal filter coefficients from \eqref{eq:c_opt} into \eqref{eq:amb_est}, the estimated Ambisonics signals are obtained. To ensure the successful encoding of all target channels, the microphone array must satisfy the following dimensionality constraint \cite{rafaely2015fundamentals}:
\begin{equation}
    (N_a + 1)^2 \leq M,
\end{equation}
which implies that the number of microphones $M$ must be at least equal to the total number of Ambisonics coefficients.

It has been demonstrated in \cite{gayer2024ambisonics} that the accuracy of Ambisonics channel encoding is highly dependent on the microphone array geometry and its corresponding steering matrix, which may lead to suboptimal reconstruction of certain channels.

\subsection{Speech Enhancement Model}
\label{ssec:speech enhancement}
The goal in a speech enhancement task is to extract a target speaker from a mixture of interfering sources and reverberations. For this task, it is usually convenient to model the time-domain noisy signal at the $m$-th microphone, $x_m(t) \in \mathbb{R}$, as:
\begin{equation}
    x_m(t) = y_m(t) + u_m(t),
\end{equation}
where $y_m(t)$ represents the clean speech of the target speaker $s(t)$ as captured by the $m$-th microphone, and $u_m(t)$ denotes the noise component at the $m$-th microphone, encompassing interfering speakers, reverberation, and ambient noise.

To extract the target speaker, the multichannel microphone signals $\mathbf{x}(t)$ are processed by a DNN to estimate the enhanced speech $\hat{s}(t)$. The performance of the DNN is typically evaluated against the clean target speech captured at the reference channel, denoted by $x_r(t)$. Here $r$ is the index of the reference microphone, which is usually the one positioned closest to the target speaker.

\section{AmbiDrop Framework}
\label{sec:ambidrop_framework}
AmbiDrop \cite{tatarjitzky2026ambidrop} is a framework designed for array-agnostic speech enhancement that utilizes Ambisonics representations as DNN inputs rather than raw microphone signals. By capturing the underlying sound field and source characteristics independently of a specific sensor geometry (as shown in Eq. (\ref{eq:amb})), Ambisonics provide a universal spatial format for training.

While ideal Ambisonics are used during training, perfect representations are unattainable during inference due to the practical limitations of accurately estimating Ambisonics signals from real-world recordings. AmbiDrop explicitly addresses this domain mismatch by incorporating a channel-wise spatial dropout mechanism during training. This prepares the model to remain robust even when the estimated Ambisonics components are degraded, distorted, or incomplete during real-world deployment.

Since the AmbiDrop framework is comprised of three core components—the DNN architecture, the Ambisonics input representation, and a dropout layer—it is expected to be applicable across a wide range of DNN models.

\subsection{Training Phase}
\label{ssec:training}
Training a DNN directly on raw microphone signals typically requires datasets from multiple array geometries, as spatial relationships vary significantly across different configurations. In contrast, Ambisonics signals maintain a fixed number of channels for a given order and consistent spatial relationships regardless of the original sensor layout. This simplifies the training process, as the network learns spatial cues from a uniform representation. 

At the training stage, as illustrated in Fig. \ref{fig:processing_chain_train}, the DNN is fed with ideal Ambisonics signals $\mathbf{a}_{nm}(t)$ up to order $N_a$. The input first passes through a channel-wise dropout layer, which simulates potential estimation failures at the inference stage as will be explained later in Sec. \ref{ssec:dropout}. The resulting signals are then processed by the DNN. For loss calculation, the omnidirectional harmonic $a_{00}(t)$ is utilized as the reference signal due to its spatial uniformity.

\subsection{Inference Phase}
\label{ssec:inference}
During inference, as illustrated in Fig. \ref{fig:processing_chain_inference}, Ambisonics signals are estimated from microphone signals of an arbitrary array using ASM \cite{gayer2024ambisonics}. These estimated signals are then processed by the DNN, bypassing the dropout layer to allow the model to leverage all available spatial information from the estimated channels.

\subsection{Dropout Layer}
\label{ssec:dropout}
Estimation methods such as ASM often introduce channel-specific errors depending on the array geometry \cite{gayer2024ambisonics}. Specifically, poorly estimated channels tend to exhibit low amplitudes, approaching zero in frequency bins where reconstruction is ill-posed. This physical behavior mirrors the dropout mechanism used in deep learning. By applying a channel-wise dropout layer to ideal Ambisonics during training, we simulate these reconstruction artifacts. This encourages the DNN to perform well even when specific channels are missing or corrupted, allowing it to maintain high performance despite the domain mismatch caused by imperfect Ambisonics encoding during inference.

\section{Baseline DNN architectures}
\label{sec:baseline}
This section describes the DNN architectures evaluated in this study. Each architecture is employed in two configurations: first, as a standalone geometry-dependent baseline (e.g. ``DNN”); and second, as the backbone for the proposed array-agnostic framework (referred to as ``DNN+AmbiDrop'').

\subsection{FT-JNF}
\label{ssec:ftjnf}
FT-JNF \cite{tesch2022insights} is a multichannel DNN designed as a joint non-linear filter that integrates spatial, spectral, and temporal cues. Operating in the Short-Time Fourier Transform (STFT) domain, the network is optimized to estimate a complex Ideal Ratio Mask (cIRM).

The multichannel time-domain signals $\mathbf{x}(t) \in \mathbb{R}^M$ are transformed into the STFT domain, where $X_m(t,f)$ denotes the coefficient for the $m$-th microphone at time frame $t$ and frequency bin $f$. The resulting input tensor $\mathbf{X} \in \mathbb{C}^{M \times T \times F}$ is formed such that $\mathbf{X}(t,f) = [X_1(t,f), \dots, X_M(t,f)]^T$. The network processes the concatenated real and imaginary components through two Bidirectional Long Short-Term Memory (BLSTM) layers followed by a linear projection. The enhanced spectrum $\hat{S}(t,f)$ is obtained by applying the estimated complex mask $M(t,f)$ to the reference channel $X_r(t,f)$:
\begin{equation} 
\hat{S}(t,f) = M(t,f) \cdot X_r(t,f). \label{eq:net_output} 
\end{equation} 
The time-domain enhanced signal $\hat{s}(t)$ is reconstructed via the inverse STFT (iSTFT). In the \textit{FT-JNF+AmbiDrop} configuration, the microphone signals $\mathbf{x}(t)$ are replaced by the complex-valued Ambisonics signals $\mathbf{a}_{nm}(t)$. Since both representations share the same complex STFT structure, the underlying network architecture remains compatible without modification.

\subsection{IC Conv-TasNet}
\label{ssec:convtasnet}
The Inter-Channel Convolutional Time-domain Audio Separation Network (IC Conv-TasNet) \cite{lee2111inter} is a time-domain masking-based framework utilizing an encoder-separator-decoder architecture. This model explicitly exploits inter-channel cues and spatial diversity to isolate the target source.

The multichannel time-domain signals $\mathbf{x}(t) \in \mathbb{R}^M$ are first mapped into a high-dimensional latent space by a linear encoder. A separator, consisting of stacked dilated Temporal Convolutional Networks (TCNs), estimates a real-valued mask $\in [0, 1]$ based on the fused multichannel latent representation. This mask is element-wise multiplied with the encoded representation of the reference channel $x_r(t)$. Finally, a linear decoder reconstructs the enhanced time-domain signal $\hat{s}(t)$. For the \textit{IC Conv-TasNet+AmbiDrop} variant, the network is modified to accept real-valued Ambisonics signals, $\mathbf{a}_{nm,\text{real}}(t)$, as the primary input.

\section{Simulation Study}
\label{sec:sim_experiment}
This section first details the simulated datasets and DNN configurations employed to train and evaluate the AmbiDrop framework. Then, focusing on the extraction of a target speaker from a multi-talker mixture, we assess performance within a controlled acoustic environment. The DNN models described in Section~\ref{sec:baseline} serve as our baseline; we compare their standalone performance against their performance when integrated with the AmbiDrop framework to quantify the advantages of our proposed approach.

\subsection{Setup}
\label{ssec:simulated_setup}

\begin{figure*}[t] 
    \centering
    
        \begin{tabular}{cccc}
            \includegraphics[width=0.22\textwidth]{"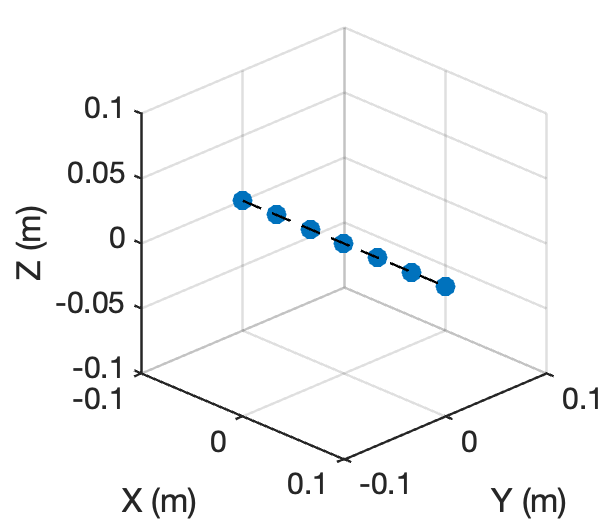"} & 
            \includegraphics[width=0.22\textwidth]{"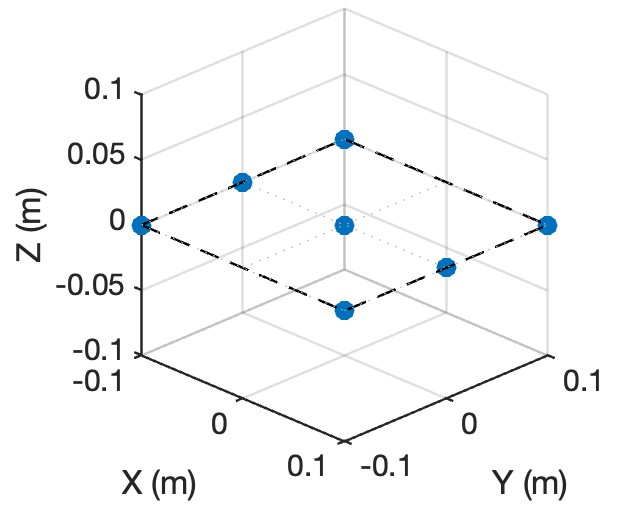"} & 
            \includegraphics[width=0.22\textwidth]{"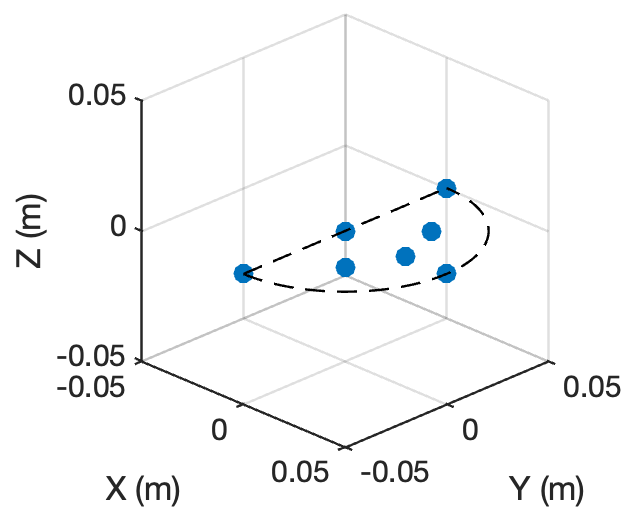"} & 
            \includegraphics[width=0.22\textwidth]{"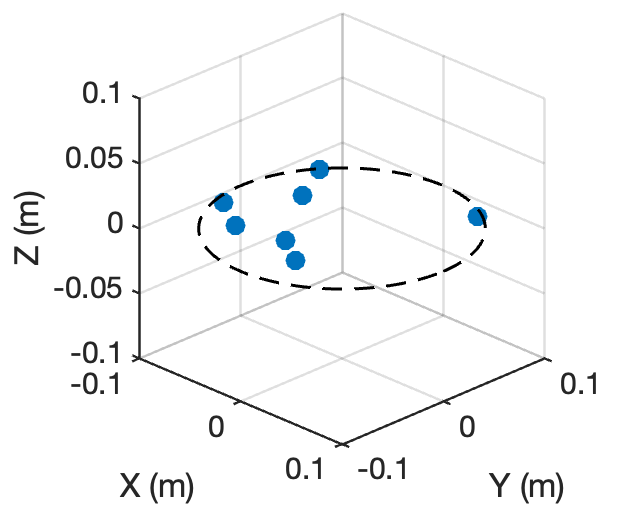"} \\
            (1) & (2) & (3) & (4) \\[2ex] 
            
            \includegraphics[width=0.22\textwidth]{"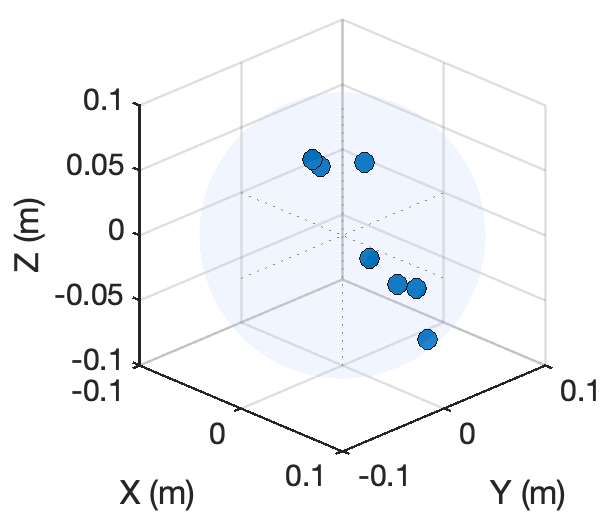"} & 
            \includegraphics[width=0.22\textwidth]{"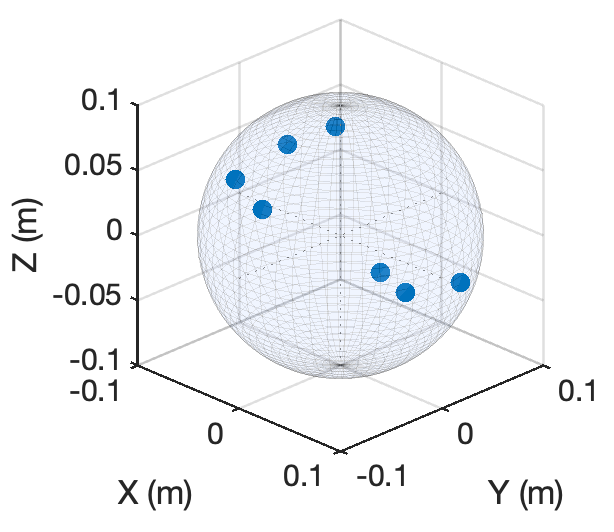"} & 
            \includegraphics[width=0.22\textwidth]{"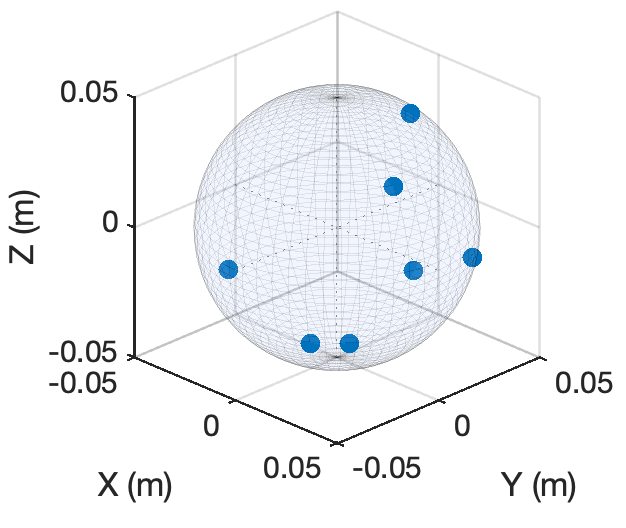"} & 
            \includegraphics[width=0.22\textwidth]{"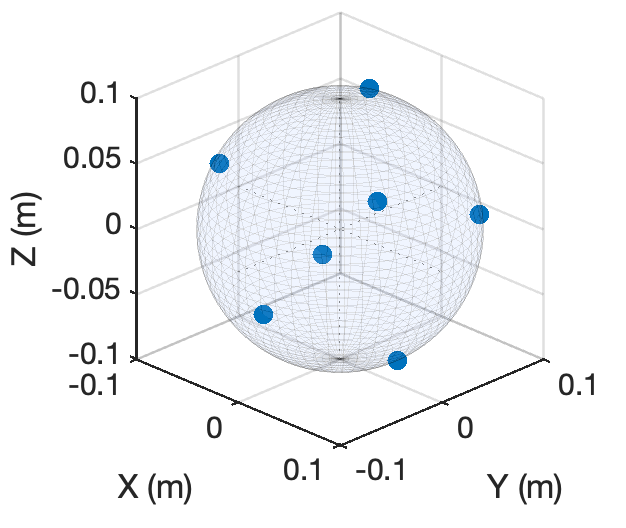"} \\
            (5) & (6) & (7) & (8) \\[2ex]
        \end{tabular}

        \begin{tabular}{cc}
            \includegraphics[width=0.22\textwidth]{"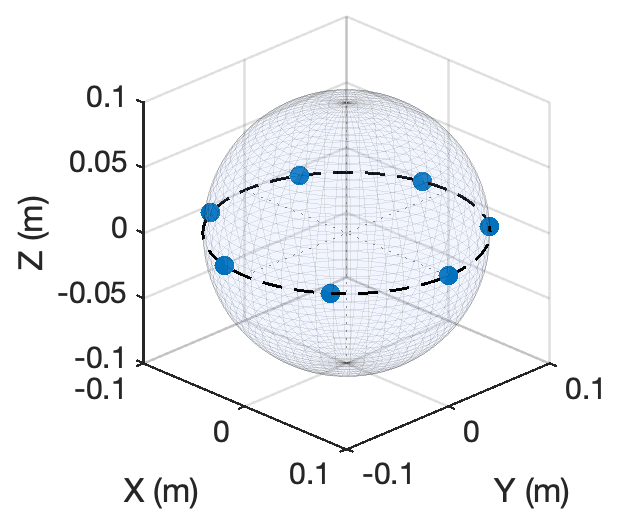"} & 
            \includegraphics[width=0.22\textwidth]{"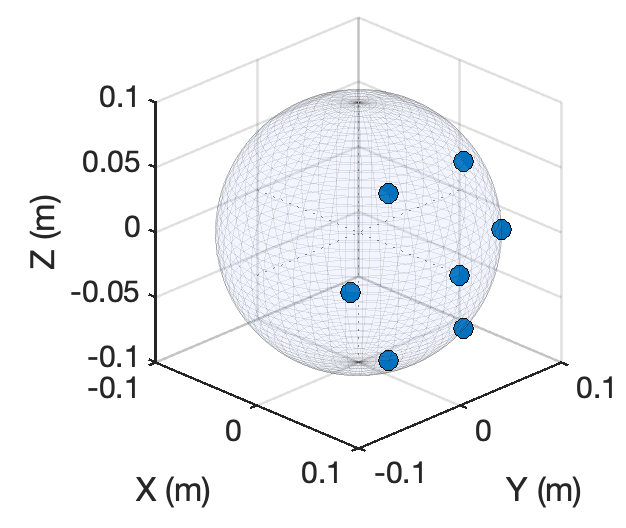"} \\
            (9) & (10) \\
        \end{tabular}
        
        \caption{\textbf{Training Arrays.} (1-4) 1D/2D free-field arrays; (5) free-field spherical volume with internal and surface-mounted microphones; (6-10) microphones mounted on rigid spheres. Blue dots mark microphone locations, while dashed lines serve as auxiliary guides to illustrate the array geometry. All arrays are bounded by spheres of radius $r=0.05$ or $0.1\,$m, except for array (2), which is defined by a $0.1\,$m planar rectangle.}
        \label{fig:train arrays}

\end{figure*}

\begin{figure*}[t] 
    \centering
    
        \begin{tabular}{cccc}
            \includegraphics[width=0.22\textwidth]{"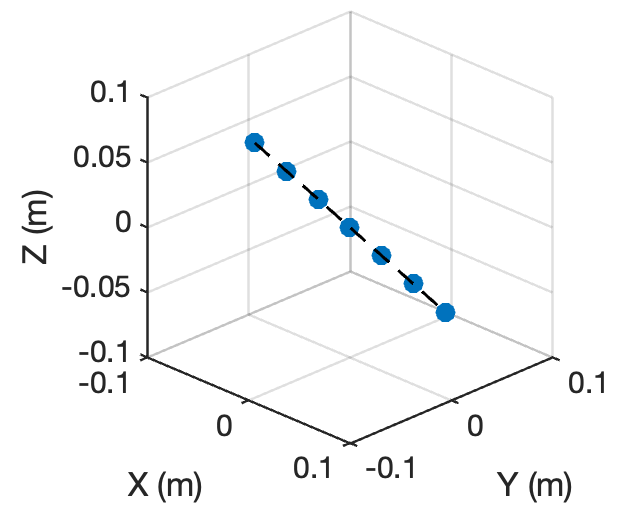"} & 
            \includegraphics[width=0.22\textwidth]{"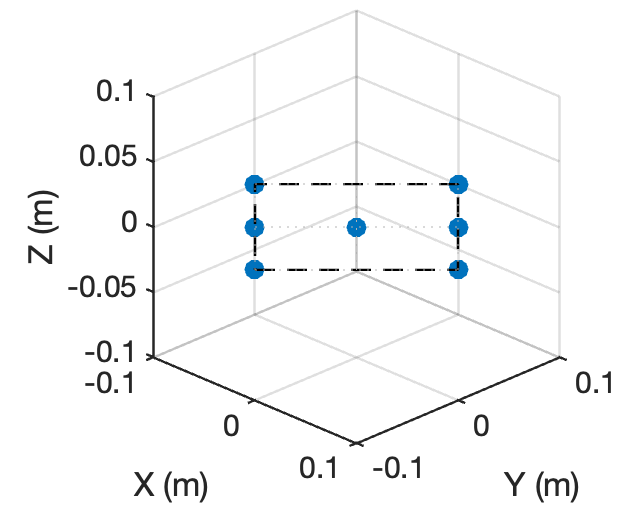"} & 
            \includegraphics[width=0.22\textwidth]{"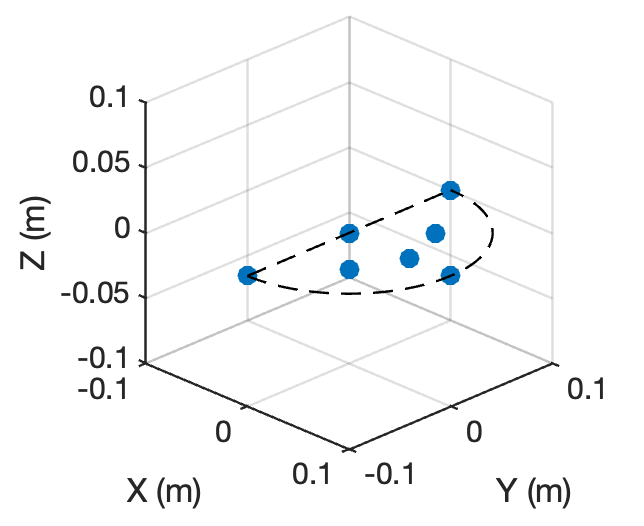"} & 
            \includegraphics[width=0.22\textwidth]{"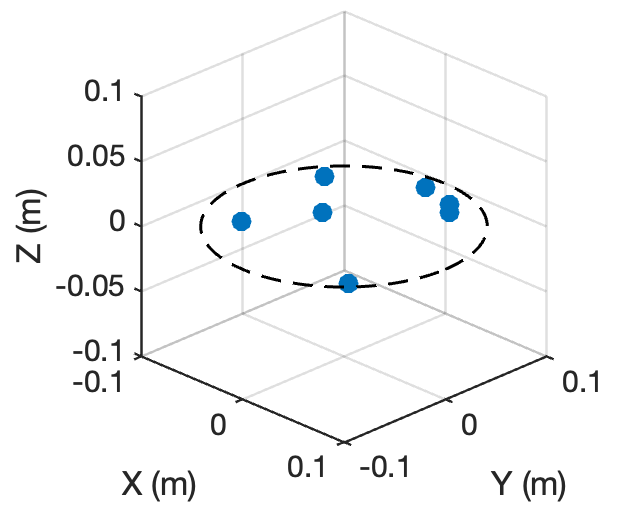"} \\
            (1) & (2) & (3) & (4) \\[2ex] 
            
            \includegraphics[width=0.22\textwidth]{"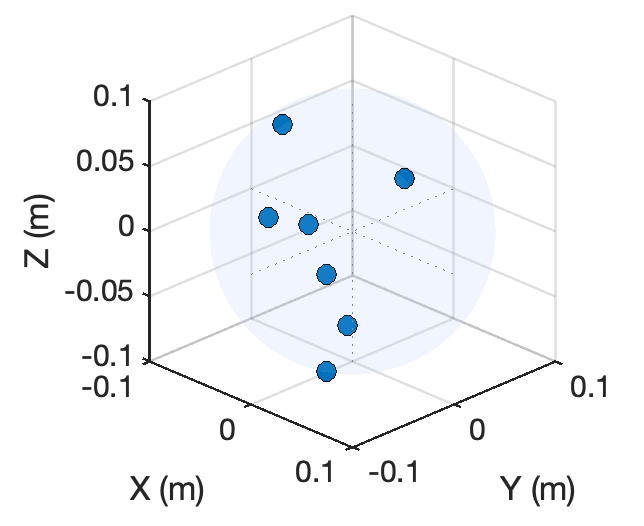"} & 
            \includegraphics[width=0.22\textwidth]{"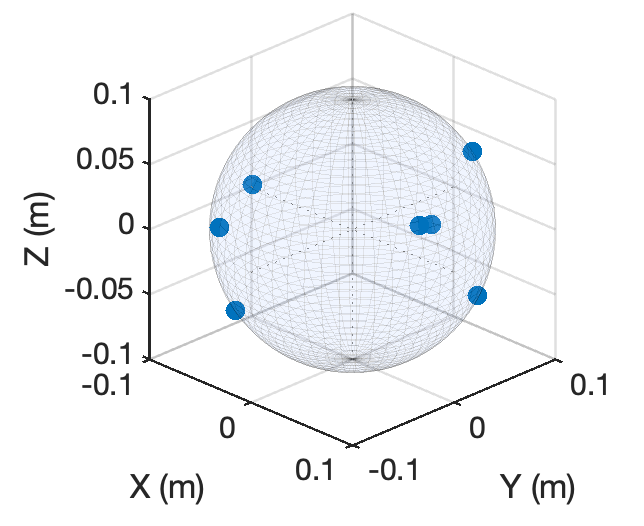"} & 
            \includegraphics[width=0.22\textwidth]{"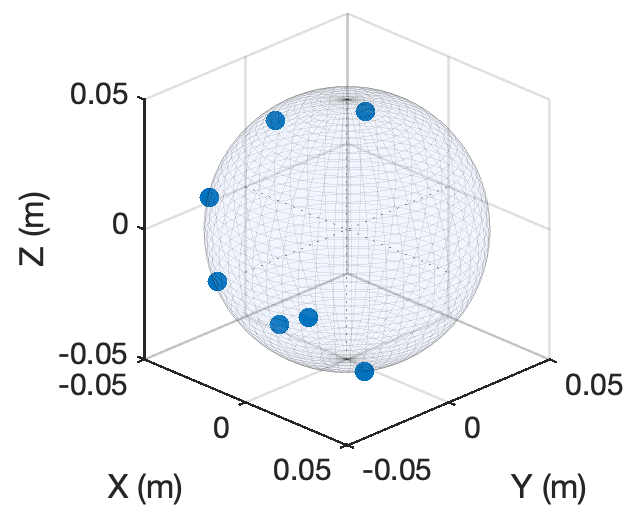"} & 
            \includegraphics[width=0.22\textwidth]{"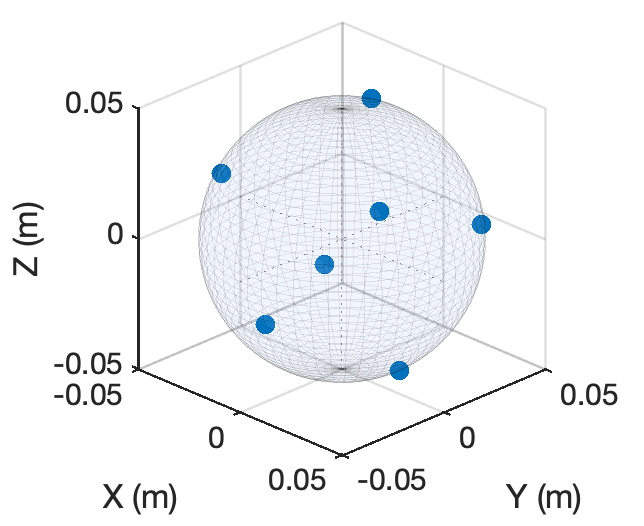"} \\
            (5) & (6) & (7) & (8) \\[2ex]
        \end{tabular}

        \begin{tabular}{cc}
            \includegraphics[width=0.22\textwidth]{"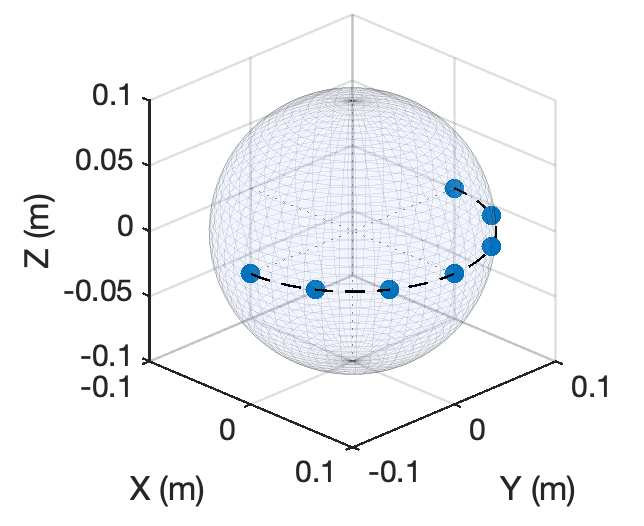"} & 
            \includegraphics[width=0.22\textwidth]{"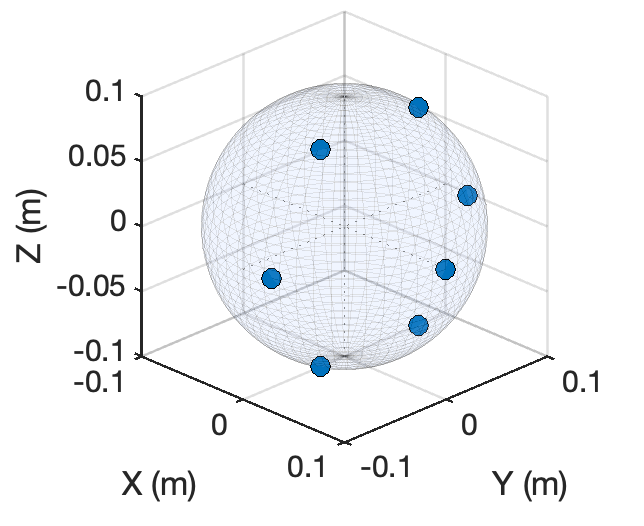"} \\
            (9) & (10) \\
        \end{tabular}
        
        \caption{\textbf{Test Arrays.} (1-4) 1D/2D free-field arrays; (5) free-field spherical volume with internal and surface-mounted microphones; (6-10) microphones mounted on rigid spheres. Blue dots mark microphone locations, while dashed lines serve as auxiliary guides to illustrate the array geometry. All arrays are bounded by spheres of radius $r=0.05$ or $0.1\,$m.}
        \label{fig:test arrays}

\end{figure*}

The simulated acoustic scenes were generated using the image method \cite{allen1979image} implemented in MATLAB. The room and speaker configurations follow the experimental setup established in \cite{tesch2022insights}, with modifications specifically to the microphone array geometries. The target speaker is positioned in the positive x-axis direction, accompanied by five competing interferers distributed throughout the acoustic scene. In this study, we generated 20 distinct 7-microphone arrays encompassing 1D, 2D, and 3D configurations—including standard geometric layouts and randomized distributions—utilizing both free-field and rigid-sphere steering functions. The resulting geometries were partitioned into two sets: training arrays (see Fig.~\ref{fig:train arrays}) and test arrays (see Fig.~\ref{fig:test arrays}), with each set comprising a unique collection of configurations. All arrays were constrained to a volume on or within a sphere of $0.1\,$m radius, except for training array (2), which is defined by a $0.1\,$m planar rectangle.

For the baseline DNN models, the training and validation datasets were synthesized using the training array configurations. This set includes a diverse range of geometries to promote generalization to unseen arrays. The clean reference signal corresponds to the microphone closest to the target speaker—positioned at the front of the array in the positive $x$-direction—and consists of the direct-path component of the target signal within the room.

In contrast, the AmbiDrop training and validation datasets were constructed using ideal Ambisonics signals up to order $N_a=2$, resulting in 9 channels. These were calculated from the direct sound and room reflections using Eq. (\ref{eq:amb}) and the same experimental setup as the baseline. For this model, the clean reference signal is defined as the direct-path component of the target speaker in the omnidirectional Ambisonics channel, $a_{00}(t)$.
To simulate realistic conditions, additive noise was introduced to both the baseline and AmbiDrop datasets, to achieve a signal-to-noise ratio (SNR) of $30\,$dB.

Two simulated inference datasets were generated to evaluate the baseline models and the AmbiDrop framework. The first set utilized the training arrays used for training the baseline models, while the second set utilized previously unseen test arrays. Each dataset contains an identical number of examples, with source and room configurations kept constant across all arrays to ensure a direct and fair comparison. 
At the inference stage, the input provided to both the baseline DNN and the DNN with AmbiDrop is identical. Consequently, the input (noisy) Scale-Invariant Signal-to-Distortion Ratio (SI-SDR) was calculated for each recording relative to the clean target speech captured by the reference microphone.

Clean speech signals were obtained from the WSJ0 dataset \cite{garofolo2007wsj0} and sampled at $16\,$kHz. For the baseline DNN training and validation sets, 1,000 and 170 recordings were simulated per array, respectively, resulting in a total of 10,000 training and 1,700 validation examples. For the AmbiDrop framework, 6,000 training and 1,000 validation examples were generated. Finally, each inference dataset comprised 300 examples per array.

\subsection{Methodology}
\label{ssec:sim_methodology}
For the training and validation datasets, 6-second recordings were utilized, with each noisy recording temporally aligned to the start of the corresponding clean target signal.

The FT-JNF network input signals were preprocessed by applying a STFT with a $32\,$ms Hamming window and 50\% overlap. The architecture consisted of two BLSTM layers, each configured with 64 units. The final linear layer produced a two-dimensional output corresponding to the real and imaginary parts of the estimated complex mask. The network, totaling 142,594 parameters, was trained for 250 epochs using the ADAM optimizer with a learning rate of 0.001, a weight decay of $10^{-5}$, and a batch size of 8. The model instance achieving the lowest validation loss was selected for evaluation.

For the IC-ConvTasNet architecture, we followed the notation and parameterization described in \cite{lee2111inter}. The encoder output dimension was set to 512, the bottleneck layer feature number to 128, and the channel dimension to 8. The model comprised a single TCN stack containing 8 TCN layers, utilizing a $3 \times 3$ kernel size for the convolutional blocks. The encoder and decoder processed time-domain signals using a 16~ms window with 50\% overlap. The network, totaling 405,100 parameters, was trained for 100 epochs using the ADAM optimizer with a learning rate of 0.001 and a batch size of 64.

For models integrated with the AmbiDrop framework, input channel-wise dropout was applied during training, randomly dropping up to 3 channels with a probability of 0.4. During the inference stage, microphone signals were first encoded to the Ambisonics domain using the ASM method (Section~\ref{ssec:Ambisonics Encoding}), assuming sensor noise with a SNR of $30\,$dB for the encoding process.

All networks were optimized using the negative SI-SDR \cite{le2019sdr} as the loss function. Performance was evaluated using three objective metrics: SI-SDR, Perceptual Evaluation of Speech Quality (PESQ), and Short-Time Objective Intelligibility (STOI).

\subsection{Results}
\label{ssec:sim_results}
\begin{table*}[!t]
\centering
\fontsize{9}{12}\selectfont
\setlength{\tabcolsep}{8pt}
\renewcommand{\arraystretch}{1.2}
\begin{tabular}{|c|l|c|c|c|}
\hline
\multicolumn{1}{|c|}{\textbf{Dataset}} & \multicolumn{1}{c|}{\textbf{Method}} & \textbf{SI-SDR} (dB) $\uparrow$ & \textbf{PESQ} $\uparrow$ & \textbf{STOI} $\uparrow$ \\ \hline

\multirow{5}{*}{\centering Training Arrays} 
 & Noisy (Input) & -6.2& 1.17& 0.6\\ \cline{2-5}
 & FT-JNF (Baseline) & \textbf{5.93}& 1.72& 0.85\\ 
 & FT-JNF + AmbiDrop & 5.06& \textbf{1.81}& \textbf{0.86}\\ \cline{2-5}
 & IC-ConvTasNet (Baseline) & \textbf{2.92}& 1.37& 0.78\\ 
 & IC-ConvTasNet + AmbiDrop & 2.48& \textbf{1.49}& \textbf{0.79}\\ \hline

\multirow{5}{*}{\centering Test Arrays} 
 & Noisy (Input) & -6.35& 1.17& 0.6\\ \cline{2-5}
 & FT-JNF (Baseline) & -15.08& 1.38& 0.54\\ 
 & FT-JNF + AmbiDrop & \textbf{4.77}& \textbf{1.78}& \textbf{0.84}\\ \cline{2-5}
 & IC-ConvTasNet (Baseline) & -12.27& 1.24& 0.45\\ 
 & IC-ConvTasNet + AmbiDrop & \textbf{1.4}& \textbf{1.43}& \textbf{0.76}\\ \hline

\end{tabular}
\caption{Performance across simulated datasets. For each dataset, SI-SDR, PESQ and STOI, averaged over the dataset, are presented for the baseline DNNs, and compared against the same architectures integrated with the AmbiDrop framework. Bold values indicate the best performance for each dataset.}
\label{tab:simulated_results}
\end{table*}
In this section, we evaluate the baseline DNN models, with and without the AmbiDrop framework, using the two simulated inference datasets defined in Section~\ref{ssec:simulated_setup}. The results are summarized in Table~\ref{tab:simulated_results}. All models were trained using their respective simulated training sets as described in Section~\ref{ssec:simulated_setup}.

The first experiment utilize the familiar training array geometries. While these geometries were included during the training phase of the baseline models, they remain effectively ``unseen'' for the models integrated with AmbiDrop due to the inherently array-agnostic nature of the training phase. As shown in Table~\ref{tab:simulated_results}, the baseline models slightly outperform the AmbiDrop-integrated models in terms of SI-SDR. This result is expected, given that the baselines are specifically optimized for these array configurations. Notably, however, each DNN combined with AmbiDrop maintains robust performance despite never having encountered these geometries during training, remaining highly competitive with the baseline models.

In the second experiment, utilizing the test array dataset, the geometries are unseen for both the baseline and the AmbiDrop-enhanced models. The results demonstrate that both baseline models suffer from significantly degraded performance, with enhanced SI-SDR values falling below the noisy input level. This indicates poor generalization to unseen microphone configurations. In contrast, the AmbiDrop-integrated DNNs maintain strong performance across all metrics. The moderate drop of approximately $1\,$dB observed in the test array experiment for the AmbiDrop models is likely attributed to the specific test geometries being more sensitive to ASM encoding errors rather than a lack of generalization, as all arrays are equally novel to the AmbiDrop framework.

\begin{figure}[t]
    \centering
    \includegraphics[width=\columnwidth]{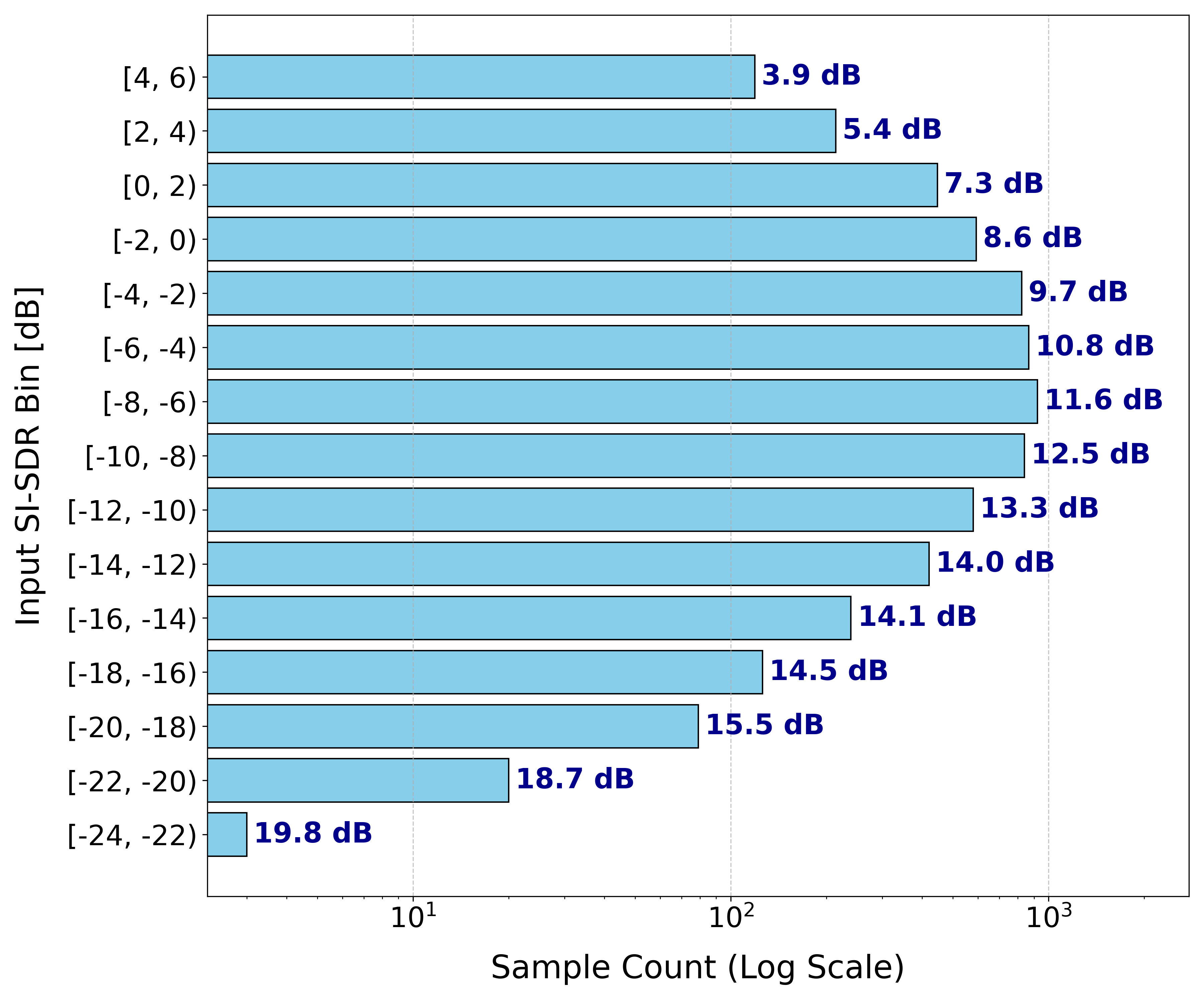} 
    \caption{Noisy SI-SDR distribution across simulated inference datasets and mean SI-SDRi per bin.}
    \label{fig:histogram}
\end{figure}

Furthermore, we provide a histogram (Fig.~\ref{fig:histogram}) of all recordings across both simulated inference datasets. The histogram illustrates the SI-SDR distribution of the noisy recordings and the frequency of recordings within each SI-SDR range. For each bin, we report the mean SI-SDR improvement (SI-SDRi) achieved by the FT-JNF + AmbiDrop model. The data suggests an inverse relationship between input quality and enhancement gain: the lower the initial SI-SDR, the greater the performance improvement achieved by the network. This can be explained by the nature of the SI-SDR measure, quantifying both noise reduction and signal distortion. At low input SI-SDR values, considerable noise reduction can be applied before distortion start to affect SI-SDR improvement. 

These results highlight the robustness and generalization capabilities of the AmbiDrop framework when applied to unseen arrays through array-agnostic training. Furthermore, the consistent performance observed across two different DNN architectures suggests that the framework may be adaptable to various network configurations. While the full extent of this compatibility remains a subject for future research, these initial findings are promising for its integration with different speech enhancement models.

\section{Experimental Study with Project Aria Glasses}
\label{sec:real_experiment}
This section aims to evaluate the proposed AmbiDrop method with real-world data, measured on an actual microphone array device - the Aria glasses.

\subsection{Setup}
\label{ssec:real_setup}

To validate the proposed framework, we conducted real-world experiments using Project Aria glasses \cite{engel2023project} (Aria Gen 1, large model). This experiment evaluates the baseline and AmbiDrop-enhanced models on physical recordings with microphone positions that mirror the simulated configurations described in Sec.~\ref{ssec:simulated_setup}, maintaining the objective of extracting a target speaker from a multi-talker mixture.

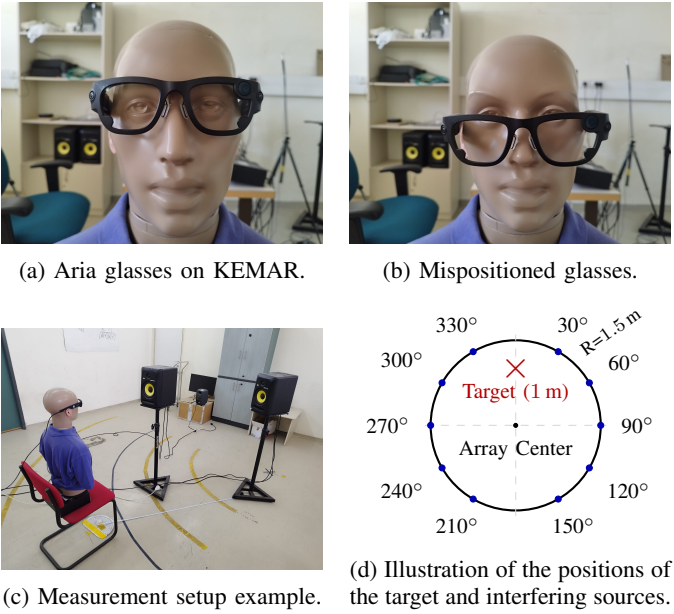
\begin{figure}[t]
    \centering
    \begin{subfigure}[b]{0.48\columnwidth}
        \centering
        \includegraphics[width=\textwidth]{{"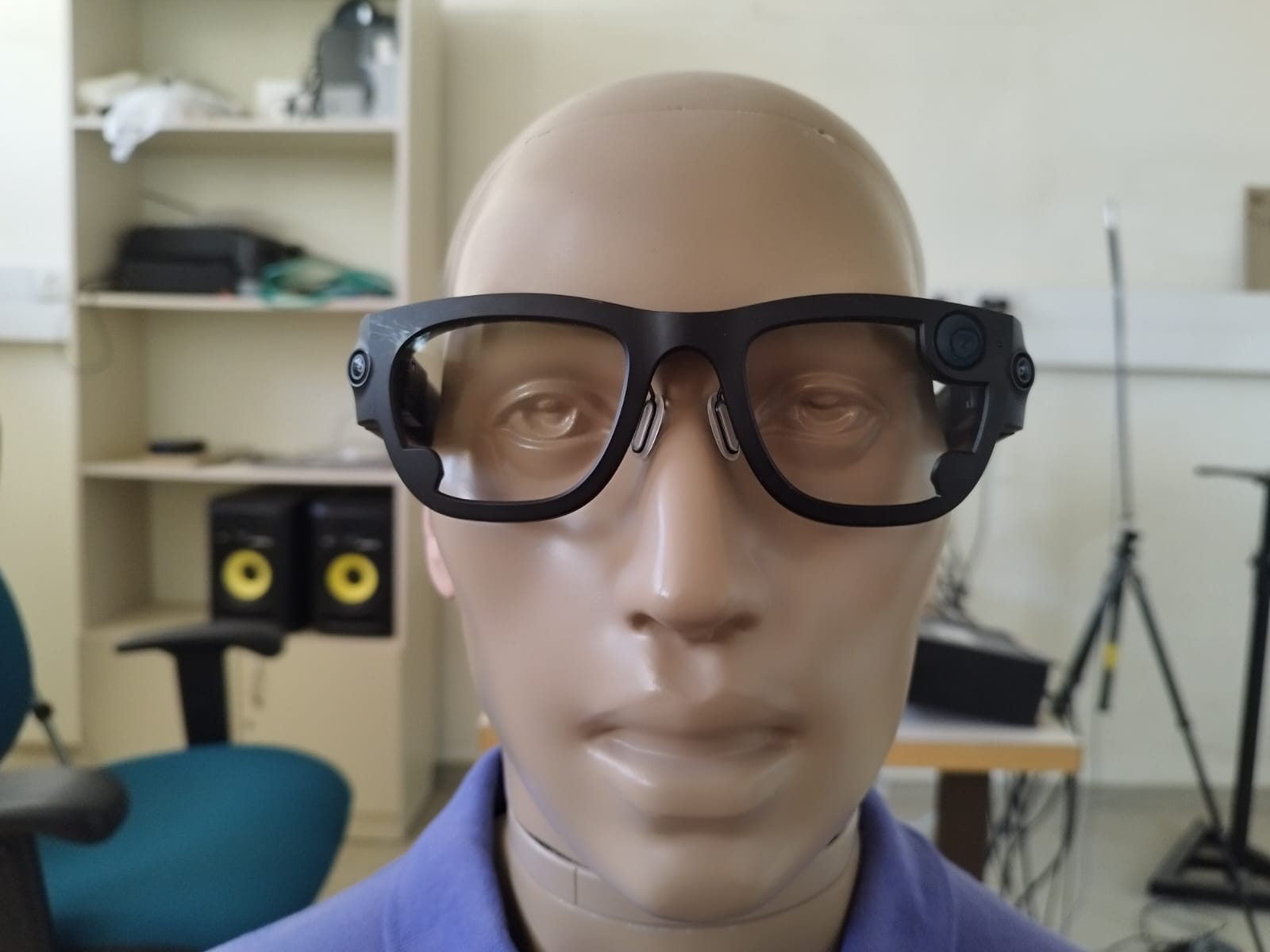"}}
        \caption{Aria glasses on KEMAR.}
        \label{fig:correct_aria}
    \end{subfigure}
    \hfill
    \begin{subfigure}[b]{0.48\columnwidth}
        \centering
        \includegraphics[width=\textwidth]{{"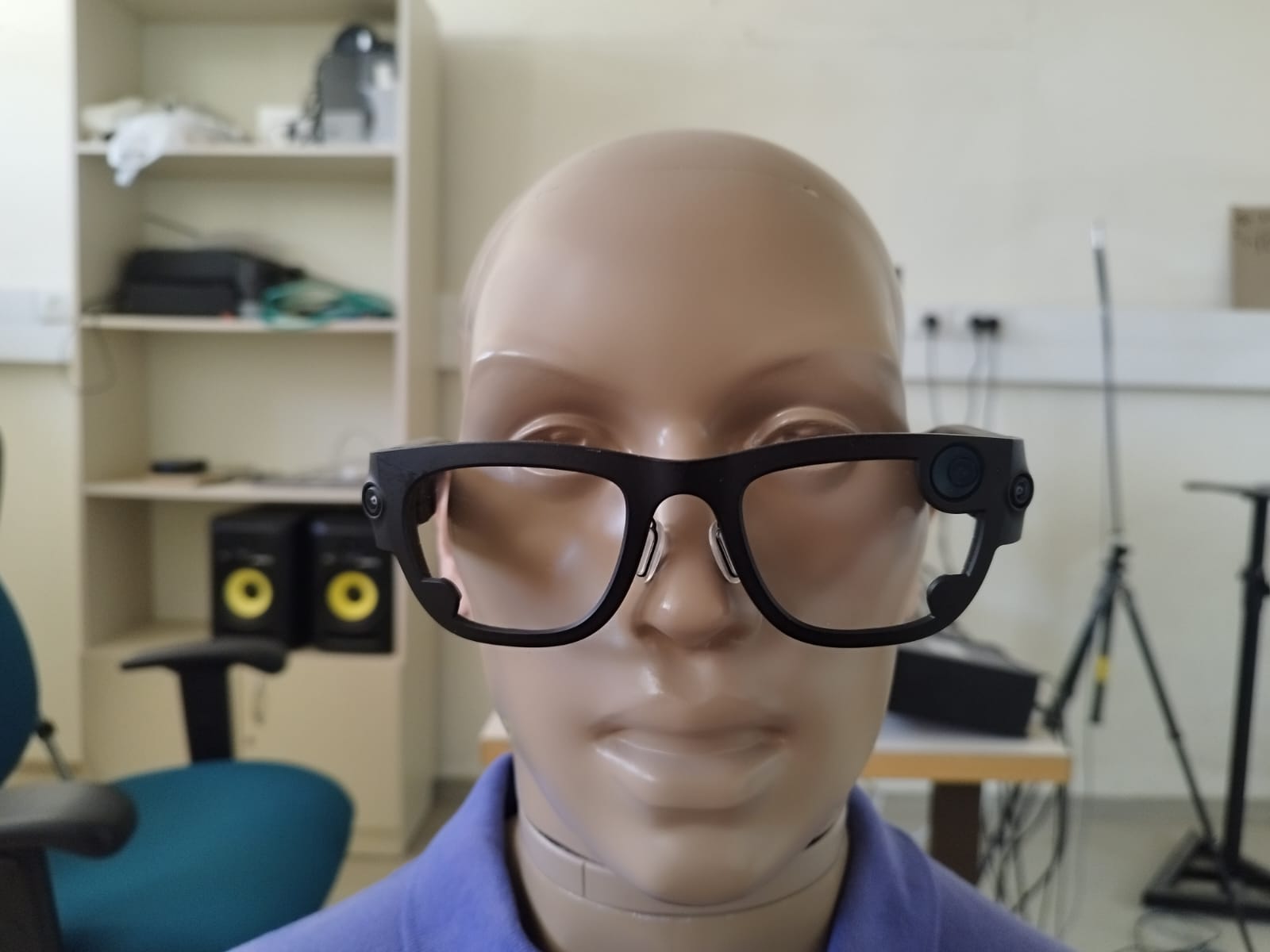"}} 
        \caption{Mispositioned glasses.}
        \label{fig:mispositioned}
    \end{subfigure}
    
    \vspace{1mm}

    \begin{subfigure}[b]{0.48\columnwidth}
        \centering
        \includegraphics[width=\textwidth]{{"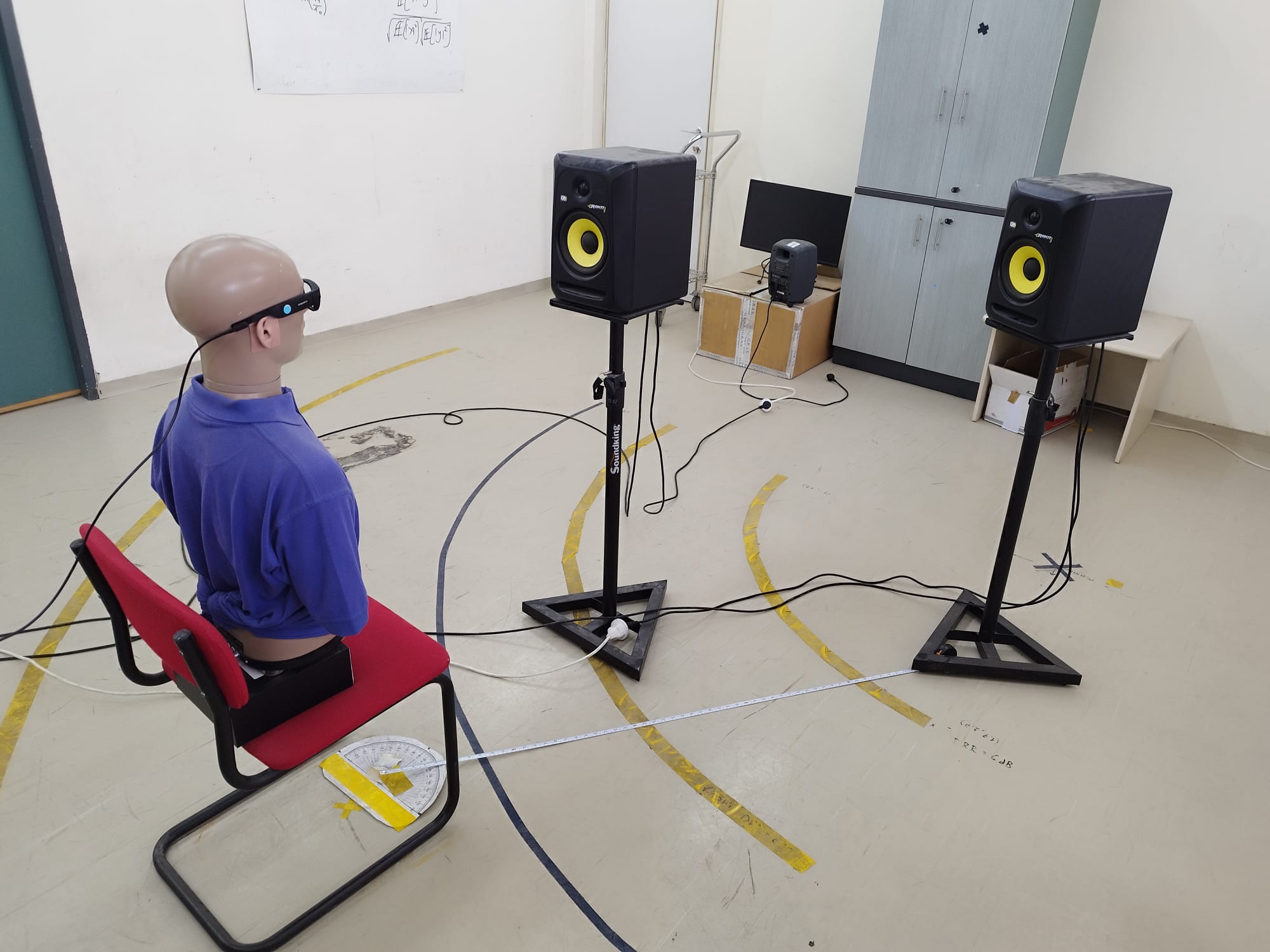"}} 
        \caption{Measurement setup example.}
        \label{fig:measurment}
    \end{subfigure}
    \hfill
    \begin{subfigure}[b]{0.48\columnwidth} 
        \centering
        \begin{tikzpicture}[scale=0.75] 
            \draw[thick] (0,0) circle (1.5); 
            \node[font=\scriptsize, anchor=south west, rotate=35] at (1.2,1.0) {R=$1.5\,$m};
            \draw[gray!30, dashed] (-1.6,0) -- (1.6,0);
            \draw[gray!30, dashed] (0,-1.6) -- (0,1.6);
            \node[font=\Large, color=red!70!black] at (0,1) {$\times$};
            \node[font=\footnotesize, anchor=north, color=red!70!black] at (0,0.9) {Target ($1\,$m)};
            \foreach \angle in {30,60,90,120,150,210,240,270,300,330}
            {
                \filldraw[blue!70!black] ({90-\angle}:1.5) circle (1.5pt);
                \node[anchor={270-\angle}, font=\footnotesize] at ({90-\angle}:1.7) {\angle$^{\circ}$};
            }
            \filldraw[black] (0,0) circle (1pt);
            \node[font=\footnotesize, anchor=north] at (0,-0.1) {Array Center};
        \end{tikzpicture}
        \caption{Illustration of the positions of the target and interfering sources.}
        \label{fig:doa-clockwise}
    \end{subfigure}

    \caption{Experimental setup showing (a) the Aria glasses on the KEMAR manikin, (b) a mispositioned Aria glasses on the KEMAR Manikin, (c) an example measurement setup showing the target loudspeaker and interference loudspeaker at $30^{\circ}$, and (d) an illustration of the position of the target source and all possible positions of the interfering sources.}
    \label{fig:experimental_setup_grid}
\end{figure}

The experiment was carried out in a laboratory room with width, length and height dimensions of $5.15 \times 6.3 \times 2.67\,$m, respectively. A KEMAR acoustic manikin \cite{burkhard1975anthropometric} wearing Aria glasses (Fig.~\ref{fig:correct_aria}) was positioned at a height of $1.2\,$m, located $2.6\,$m from the left wall and $4.1\,$m from the front wall. The array was oriented toward the shorter wall (width dimension). Target and interference signals were reproduced via KRK ROKIT6 loudspeakers with $0\,$dB gain, driven by a Scarlett 18i20 sound card. Physical coordinates for the speakers were referenced to the center of the woofer, while the array coordinates were referenced to the ``nose bridge" microphone of the Aria device.

The target speaker was positioned directly in front of the array ($0^{\circ}$) at a distance of $1\,$m. Interference sources were recorded at a distance of $1.5\,$m and at the same height as the array across 10 distinct DOAs at $30^{\circ}$ intervals, excluding $0^{\circ}$ and $180^{\circ}$ (see Fig.~\ref{fig:doa-clockwise}). An example of the physical placement for the target and an interfering source at $30^{\circ}$ is shown in Fig.~\ref{fig:measurment}.

The clean speech signals were taken from the same set as for the simulated study (Section~\ref{ssec:simulated_setup}). For each DOA, including the target position, multiple 10-second segments of clean speech were recorded individually to allow for control of noisy SI-SDR in post-processing. Speech signal were different for each speech source within a single recorded scene, ensuring no overlap in speech content.

Noisy mixtures were synthesized by randomly selecting five interference recordings from different DOAs and combining them with a target speaker recording. To simulate various distances and relative source levels, interference signals were scaled by random gains sampled uniformly from the range $[0.2, 0.7]$. Following this procedure, an evaluation dataset of 300 unique mixture recordings was generated.

Furthermore, to evaluate robustness against variations in glasses fittings on the head, we recorded the target and interference sources using an array that was intentionally mispositioned on the manikin's head (see Fig.~\ref{fig:mispositioned}). A second set of 300 evaluation examples was generated using these recordings in the same manner as the primary dataset.

\subsection{Methodology}
\label{ssec:real_methodology}
For the measured data evaluation, we employ the FT-JNF network with the parameters defined in Sec.~\ref{ssec:sim_methodology}. Notably, we utilize the exact baseline and AmbiDrop-enhanced models trained previously on simulated data (Sec.~\ref{ssec:simulated_setup}) to assess their zero-shot generalization capabilities. For the baseline model, the ``lower-lens left" microphone of the Aria glasses was selected as the reference channel, as it is the closest microphone to the target speaker.

Ambisonic encoding is performed using two distinct Aria ATFs: (1) a simulated ATF modeled as a rigid sphere in MATLAB based on the Aria geometry from the CHiME-8 challenge \cite{vzmolikova2024chime}, and (2) the measured ATF provided by the same challenge. To maintain consistency with the $16\,$kHz sampling rate of our clean signals and simulated ATFs, the $48\,$kHz Aria recordings and measured ATFs are downsampled to $16\,$kHz.

As described previously, two datasets were collected: one with the Aria glasses correctly positioned on a manikin head, and another with a deliberate mispositioning, both utilizing the two aforementioned ATFs. For comparison, we also included a simulated dataset with the same acoustic scenario as described in the simulation study (see Section~\ref{ssec:simulated_setup}), containing an equivalent number of recordings generated with the simulated Aria ATF.

A significant challenge in real-world recording is the variable latency between the record command and the actual start of capture, which complicates the acquisition of time-aligned reference signals. To address this, we computed the performance metrics by performing a temporal grid search, by shifting the dry clean target speech relative to the enhanced output with the aim of finding the alignment that maximizes the SI-SDR. All subsequent metrics (PESQ, STOI) are then calculated using this optimally shifted reference.

\subsection{Results}
\label{ssec:Aria_results}

\begin{table*}[!t]
\centering
\fontsize{9}{12}\selectfont
\setlength{\tabcolsep}{8pt}
\renewcommand{\arraystretch}{1.2}
\begin{tabular}{|>{\centering\arraybackslash}m{5.5cm}|l|c|c|c|}
\hline
\textbf{Dataset} & \multicolumn{1}{c|}{\textbf{Method}} & \textbf{SI-SDRi} (dB) $\uparrow$& \textbf{PESQ} $\uparrow$ & \textbf{STOI} $\uparrow$ \\ \hline

\multirow{3}{*}{Simulated Data} 
 & Noisy&  -5&  1.17&  0.63\\ \cline{2-5}
 & Baseline &  -8.87&  1.09&  0.51\\ 
 & AmbiDrop + Simulated ATF &  \textbf{9.6}& \textbf{1.81}&  \textbf{0.87}\\ \hline

\multirow{4}{*}{Measured Data - Normal Glasses Position} 
 & Noisy&  -6.74&  1.23&  0.69\\ \cline{2-5}
 & Baseline &  -7.87&  1.09&  0.5\\ 
 & AmbiDrop + Simulated ATF &  \textbf{7.34}&  1.62&  0.78\\ 
 & AmbiDrop + Measured ATF &  5.79&  \textbf{1.65}&  \textbf{0.79}\\ \hline

\multirow{4}{*}{Measured Data - Glasses Mispositioned} 
 & Noisy&  -7.1&  1.22&  0.67\\ \cline{2-5}
 & Baseline &  -9.14&  1.11&  0.45\\ 
 & AmbiDrop + Simulated ATF &  \textbf{5.07}&  1.45&  0.72\\ 
 & AmbiDrop + Measured ATF &  3.64&  \textbf{1.49}&  \textbf{0.75}\\ \hline

\end{tabular}
\caption{Performance evaluation in terms of SI-SDRi, PESQ and STOI, for (i) the simulated dataset with the baseline method and AmbiDrop with the simulated ATF, (ii) the measured dataset with normal glasses position and with the baseline method and AmbiDrop incorporating both simulated and measured ATFs, and (iii) the same as (ii) but for the measured dataset with glasses mispositioned.}
\label{tab:real_results}
\end{table*}

This section presents the performance of AmbiDrop on real-world recordings from the Aria glasses, as described in Section~\ref{ssec:real_setup}. 

\subsubsection{Simulated vs. measured datasets}
The results are presented in Table~\ref{tab:real_results}. Notably, the baseline model exhibits poor performance across all datasets, as the Aria array geometry was not encountered during its training. For the correctly positioned real-world dataset, AmbiDrop with simulated ATF encoding achieves approximately $2\,$dB less improvement than on the fully simulated dataset. This performance gap is expected, as real-world recordings involve complex noise fields and reflections that differ from the simulated environments used during training. However, as shown later in Section~\ref{ssec:dropout_ablation}, optimized dropout configurations can effectively mitigate this gap. 

\subsubsection{Simulated vs. measured ATF}
Interestingly, while the simulated ATF introduces a modeling mismatch relative to the physical device, it still outperforms the measured ATF in terms of SI-SDR. We attribute the degraded performance to probable errors incorporated in the physical measurements of the measured ATF - leading to a mismatch that exceeds the modeling imperfections found in the simulated ATF.

\subsubsection{Mispositioned Aria glasses}
In the mispositioned array experiment, we observe an additional $2\,$dB performance drop for AmbiDrop across both ATF types. The mispositioning introduced deviations in microphone positions relative to the manikin head, which were not captured by either the rigid-sphere model or the measured ATF. These findings underscore the importance of accurate array positioning, confirming that the performance of the AmbiDrop framework is dependent on the accuracy of the ATF modeling. 

In conclusion, this experiment validated the effectiveness of the AmbiDrop framework using a real-world array in a physical acoustic environments. The successful transition from simulated training to real-world data provides a robust foundation for the framework, demonstrating that its performance extends beyond idealized simulations to practical, real-world applications.

\section{Ablation Study}
\label{sec:ablation}
In this section, we conduct a series of ablation studies to evaluate the impact of various system components and hyperparameters on performance. The study is performed using the FT-JNF architecture, and evaluation is presented in terms of SI-SDR, as this metric directly aligns with the training objective of the models.

\subsection{Impact of the Dropout Layer}
\label{ssec:dropout_ablation}

\begin{figure*}[t]
    \centering
    \includegraphics[width=1.7\columnwidth]{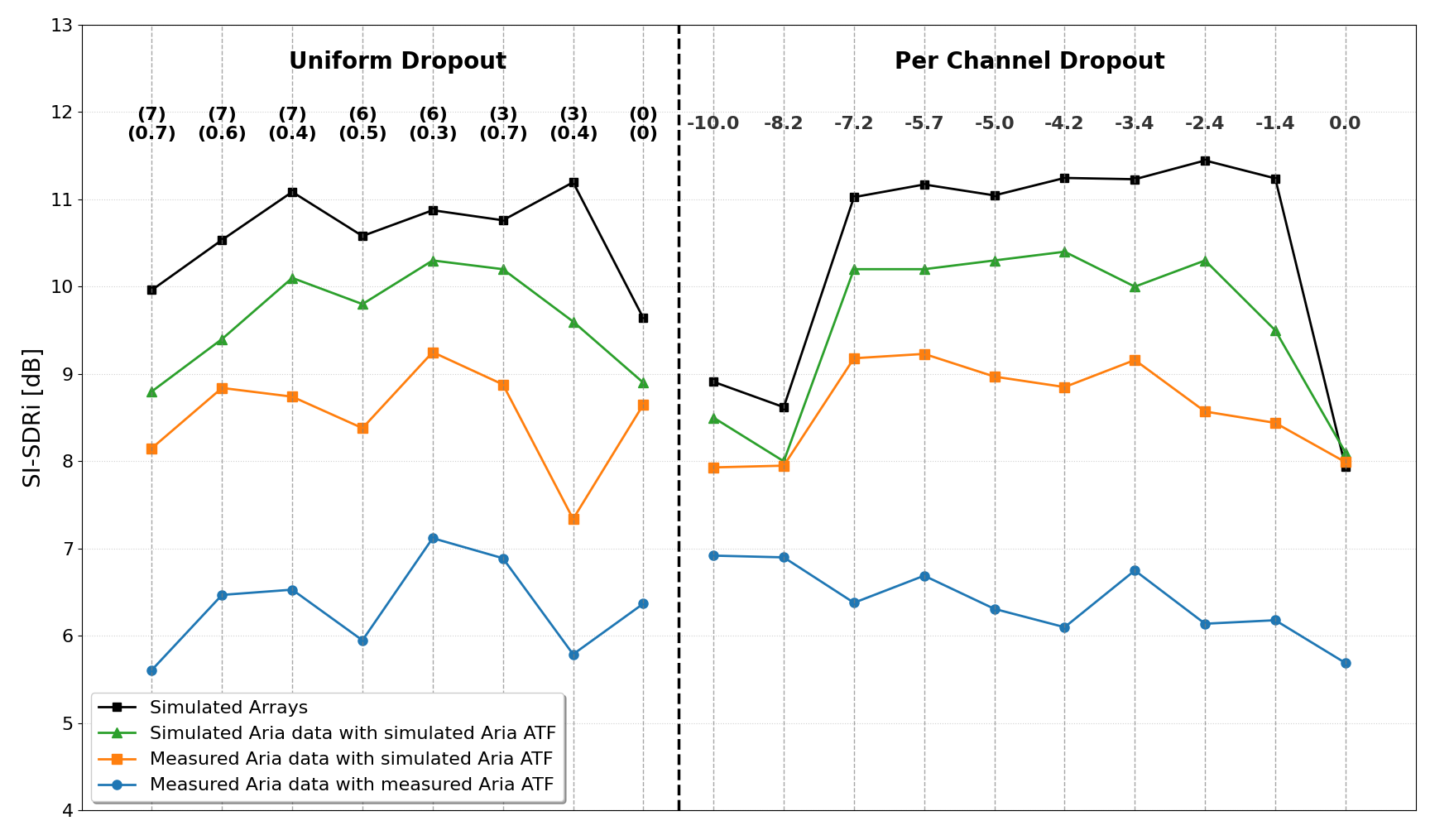}
    \caption{SI-SDRi across various dropout configurations for simulated arrays (black), simulated Aria data with simulated ATF (green), measured Aria data (normal position) with simulated and measured ATFs (orange and blue, respectively). The left section evaluates uniform dropout where each configuration is denoted by the two values of (Max. Dropped Channels) and (Dropout Probability). The right section evaluates per-channel dropout where each configuration is denoted by an error threshold.}
    \label{fig:dropout_simulated}
\end{figure*}

We investigate the influence of various input dropout alternatives. Recall that as discussed in Section~\ref{ssec:dropout}, dropout was introduced to model encoding errors during training to enhance robustness during inference. We evaluate two distinct dropout strategies:
\begin{enumerate}
    \item Uniform: A single dropout probability is applied uniformly across all channels, subject to a maximum number of dropped channels.
    \item Per Channel: Each channel is assigned a unique dropout probability derived from its specific ASM encoding errors characteristics across the dataset. No limit to the number of dropped channels is imposed.
\end{enumerate}
It is important to note that for each strategy, the $a_{00}$ channel is assigned a zero probability of being dropped, as it serves as the reference channel for the DNN models.

With the aim of supporting the computation of the unifrom and per-channel dropout probabilities, we define a broadband ASM encoding error based on Eq. (\ref{eq:amb_error}), between ideal Ambisonics ($a_{nm}$) and ASM encoded Ambisonics ($\hat{a}_{nm}$) for each channel, as follows:
\begin{equation}
    \bar{\varepsilon}_{\text{Amb}}^{nm} = \frac{\frac{1}{K} \sum_{k=1}^{K} E\left[ | \hat{a}_{nm}(k) - a_{nm}(k) |^2 \right]}{\frac{1}{K} \sum_{k=1}^{K} E\left[ | a_{nm}(k) |^2 \right]}.
    \label{eq:broadband_amb}
\end{equation} 
where $K$ is the number of frequency bins, and the expectation operator, $E[\cdot]$, is computed in practice by averaging across $32\,$ms time frames and over the 300 recordings within the inference set of each respective array. Unlike Eq. (\ref{eq:amb_error}), the error in Eq. (\ref{eq:broadband_amb}) is normalized by the broadband energy. This prevents the metric from being dominated by high-error, low-variance frequency bins, which would occur if the per-frequency ratios in Eq. (\ref{eq:amb_error}) were simply averaged.
$\bar{\varepsilon}_{\text{Amb}}^{nm}$ is calculated for each channel and for every simulated array configuration defined in Section~\ref{ssec:simulated_setup}.

For the uniform dropout strategy, the dropout probability is determined by the percentage of ``bad" channels across the entire array dataset. A channel is classified as ``bad" if its error $\bar{\varepsilon}_{\text{Amb}}^{nm}$ exceeds a specified error threshold, where we examined threshold values ranging from $-15\,$dB to $0\,$dB. For a given threshold, the maximum number of dropped channels is defined as the peak count of bad channels observed within any single array, capped at 7 to ensure that at least one non-reference channel remains active. This approach yields eight distinct configurations—three based on the defined thresholds and five selected heuristically—as summarized in Table~\ref{tab:uniform}.

Conversely, for the per-channel dropout strategy, the dropout probability is calculated individually for each channel as the percentage of arrays for which this channel is classified as ``bad". Evaluating this across the same range of error thresholds yields a unique dropout probability vector for each threshold. We evaluate ten distinct thresholds for this strategy, resulting in the parameters summarized in Table~\ref{tab:per_channel}.

\begin{table}[t]
\centering
\renewcommand{\arraystretch}{1.2}
\setlength{\tabcolsep}{4pt}
\fontsize{8}{10}\selectfont
\begin{tabular}{|c|c|c|}
\hline
\rowcolor[HTML]{EFEFEF} 
\textbf{Threshold [dB]} & \textbf{Max. Dropped Channels} & \textbf{Dropout Probability} \\ \hline
-6.7 & 7 & 0.7 \\ \hline
-4.7 & 7 & 0.6 \\ \hline
—& 7 & 0.4 \\ \hline
-2.8 & 6 & 0.5 \\ \hline
—& 6 & 0.3 \\ \hline
—& 3 & 0.7 \\ \hline
—& 3 & 0.4 \\ \hline
—& 0 & 0 \\ \hline
\end{tabular}

\caption{Parameters for the uniform dropout strategy. Threshold-based configurations are derived from the specified threshold values; entries marked with "—" indicate heuristically selected parameters.}
\label{tab:uniform}
\end{table}

\begin{table}[t]
\centering
\renewcommand{\arraystretch}{1.2}
\setlength{\tabcolsep}{4pt}
\fontsize{8}{10}\selectfont
\begin{tabular}{|c|c|c|c|c|c|c|c|c|c|}
\hline
\rowcolor[HTML]{EFEFEF} 
\textbf{Threshold [dB]} 
& \multicolumn{9}{c|}{\textbf{Dropout Probability (Channels 1-9)}} \\ \hline
& \textbf{1} & \textbf{2} & \textbf{3} 
& \textbf{4} & \textbf{5} & \textbf{6} 
& \textbf{7} & \textbf{8} & \textbf{9} \\ \hline
-10.0 & 0 & 0.55 & 0.9  & 0.6  & 1    & 1    & 1    & 1    & 1    \\ \hline
-8.2  & 0 & 0.35 & 0.75 & 0.2  & 1    & 1    & 1    & 1    & 1    \\ \hline
-7.2  & 0 & 0.15 & 0.6  & 0.2  & 0.95 & 1    & 0.95 & 1    & 0.95 \\ \hline
-5.7  & 0 & 0.15 & 0.55 & 0.15 & 0.8  & 1    & 0.95 & 1    & 0.85 \\ \hline
-5.0  & 0 & 0.1  & 0.45 & 0.15 & 0.7  & 1    & 0.85 & 1    & 0.65 \\ \hline
-4.2  & 0 & 0.1  & 0.45 & 0.1  & 0.55 & 1    & 0.85 & 1    & 0.55 \\ \hline
-3.4  & 0 & 0.1  & 0.45 & 0.1  & 0.45 & 1    & 0.75 & 1    & 0.45 \\ \hline
-2.4  & 0 & 0.05 & 0.45 & 0.05 & 0.4  & 0.95 & 0.5  & 0.95 & 0.4  \\ \hline
-1.4  & 0 & 0.05 & 0.45 & 0.05 & 0.1  & 0.75 & 0.4  & 0.75 & 0.1  \\ \hline
0& 0 & 0.05 & 0.35 & 0.05 & 0    & 0.4  & 0    & 0.3  & 0    \\ \hline
\end{tabular}

\caption{Dropout probability vectors derived from each examined error threshold within the per-channel dropout strategy.}
\label{tab:per_channel}
\end{table}

The performance of the FT-JNF+AmbiDrop model across the different configurations for both strategies is presented in Fig.~\ref{fig:dropout_simulated}. For the simulated array dataset - comprising both the training and test geometries defined in Sec.~\ref{ssec:simulated_setup} (black curve) - most configurations yield an SI-SDRi of about $11\,$dB. When dropout is too aggressive (i.e. a threshold of $-10\,$dB), the model lacks important information and performance drops. Conversely, when dropout is too weak (i.e., a threshold of $0\,$dB), performance drops again, probably because encoding errors are insufficiently simulated. Notably, the baseline model without dropout $(0, 0)$ performs approximately $1.5\,$dB lower than most other configurations, leading to the conclusion that the incorporation of the dropout layer is an important factor in overall model performance, but performance is not highly sensitive to the dropout parameters.

Comparing the two dropout strategies reveals that the per-channel approach does not offer a significant advantage over the simpler uniform dropout method. While per-channel tuning allows for a more granular configuration, the uniform dropout setting of $(3, 0.4)$ used throughout this study remains a robust and effective choice.

Performance degrades as the data transitions from ideal simulated arrays (black curve) to real-world Aria data (orange and blue curves), as discussed in Section~\ref{ssec:Aria_results}. However, despite this shift in absolute SI-SDRi values, the performance trends remain consistent across all datasets. Nevertheless, fine-tuning of the dropout parameters based on the dataset and array types, does provide some improvements in performance.

These results demonstrate that the incorporation of a dropout layer is important for achieving high performance, although fine-tuning of dropout parameters may lead to some performance gains.

\subsection{Resilience to Microphone Failure}
\label{ssec:mic_count}

\begin{figure}[t]
    \centering
    \includegraphics[width=\columnwidth]{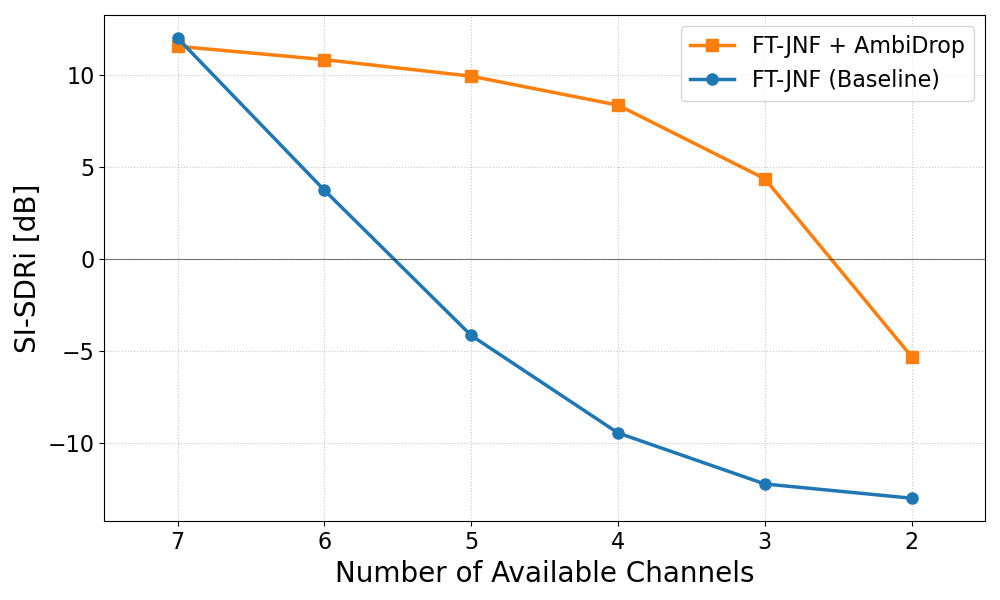}
    \caption{SI-SDRi as a function of available channels (out of the nominal 7 channels) for the FT-JNF baseline with and without AmbiDrop.}
    \label{fig:mic_ablation}
\end{figure}

Another aspect of performance with practical implications is the resilience to hardware failures in specific microphones. Because AmbiDrop training is based on the mapping of the input from a flexible array configuration to a standardized 9-channel second-order Ambisonics format, it may be interesting to investigate whether it is resilient to missing input channels.
For this investigation, we utilized an ensemble of four representative array configurations, i.e. arrays (1), (3), (5), and (8), spanning linear, planar, and spherical geometries as illustrated in Fig. \ref{fig:train arrays}. During evaluation, we randomly deactivated a specific number of microphones from the original 7-channel input, while providing the ASM stage with this information so that the deactivated channels were not included in the Ambisonics encoding process. Performance of AmbiDrop was evaluated and compared to the baseline FT-JNF method.

As shown in Fig.~\ref{fig:mic_ablation}, the baseline model's performance degrades significantly even with a single missing microphone, dropping from an initial SI-SDRi of approximately $12\,$dB down to nearly $4\,$dB. Conversely, the AmbiDrop framework maintains stable performance down to four available input channels, with the SI-SDRi only decreasing by roughly $2\,$dB, specifically from $11.5\,$dB to $9.5\,$dB. This resilience is attributed to the fact that Ambisonics channels maintain a standardized spatial structure, and the ASM can still perform effective spatial encoding provided a sufficient number of microphones remain active. When only two or three channels remain, performance degrades sharply for both models, as the spatial sampling becomes insufficient to correctly encode the Ambisonics field. 

In summary, for typical real-world sensor failures involving 1-2 missing channels, the AmbiDrop framework maintains nearly full performance, whereas the standard baseline architecture fails significantly. These results indicate that our framework effectively decouples the enhancement task from the physical integrity of the input array. 

\subsection{Network Complexity vs. Performance}
\label{ssec:net_size}

\begin{figure}[t]
    \centering
    \includegraphics[width=\columnwidth]{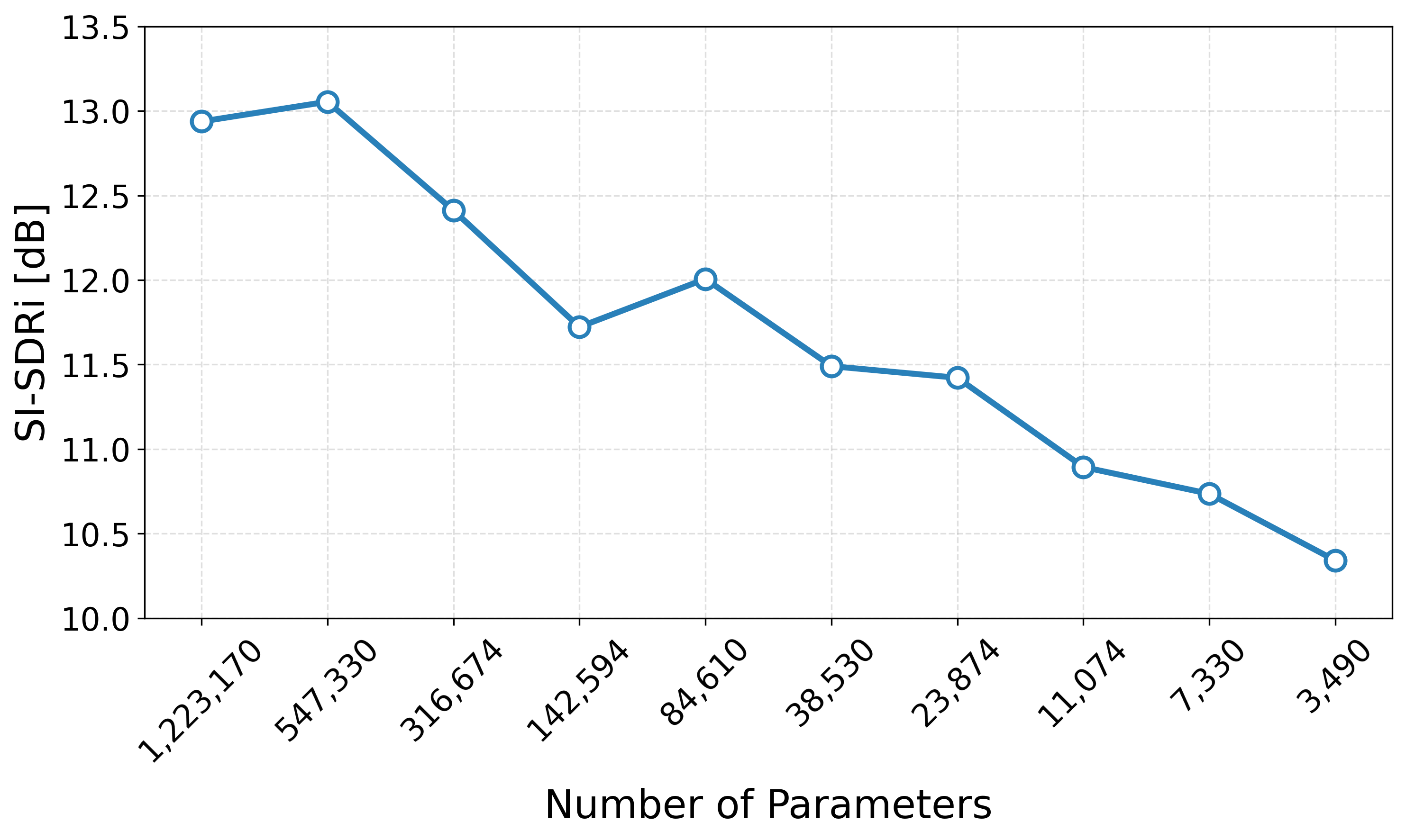}
    \caption{SI-SDRi vs. the total number of parameters of AmbiDrop's FT-JNF network.}
    \label{fig:size_ablation}
\end{figure}

We further evaluate the performance of the AmbiDrop-enhanced FT-JNF network across varying model scales. This analysis is motivated by the requirement for low-complexity architectures suitable for edge-device implementations and real-time applications.

To assess the impact of model complexity, we varied the hidden dimensions ($H_1, H_2$) of the two BLSTM layers within the FT-JNF architecture. The resulting configurations and their corresponding parameter counts are summarized in Table~\ref{tab:network_size}. Notably, the configuration $(64, 64)$ used throughout this paper consists of 142,594 parameters, representing a balanced trade-off between computational footprint and enhancement quality. 

\begin{table}[H]
    \centering
    \begin{tabular}{lc}
        \toprule
        \textbf{Network Size ($H_1, H_2$)} & \textbf{Total Parameters} \\
        \midrule
        (256, 128)& 1,223,170 \\
        (128, 128)& 547,330   \\
        (128, 64)& 316,674   \\
        (64, 64)& 142,594   \\
        (64, 32)& 84,610    \\
        (32, 32)& 38,530    \\
        (32, 16)& 23,874    \\
        (16, 16)& 11,074    \\
        (16, 8)& 7,330     \\
        (8, 8)& 3,490     \\
        \bottomrule
    \end{tabular}
    \caption{Network size and total parameter count for various model configurations, where ($H_1, H_2$) denotes the hidden dimensions of the two BLSTM layers in the FT-JNF network.}
    \label{tab:network_size}
\end{table}

The mean SI-SDRi results across the simulated dataset are illustrated in Fig.~\ref{fig:size_ablation}. Our findings indicate that reducing the model complexity by two orders of magnitude—from approximately $1.2\,$ million parameters down to approximately 11,000—results in a performance degradation of only $2\,$dB. Even at the extreme lower bound of 3,490 parameters, the framework maintains an SI-SDRi of nearly $10.5\,$dB. 

These results demonstrate that the AmbiDrop framework is remarkably robust to model downsizing. We hypothesize that this framework, potentially integrated with even more efficient DNN architectures, is well-suited for deployment on hardware with highly constrained computational resources.

\section{Conclusion}
\label{sec:conclusion}

In this work, we presented AmbiDrop, an array-agnostic speech enhancement framework. By utilizing ideal Ambisonics as the DNN input and incorporating a dropout layer during training, AmbiDrop decouples the training process from specific array geometries. This approach allows the model to handle encoding errors when processing ASM-encoded Ambisonics during inference.

The framework demonstrated generalization across 20 simulated arrays, spanning one, two, and three dimensions using both free-field and rigid-sphere ATF configurations. This generalization was also demonstrated across several DNN architectures and measured data from Project Aria glasses. Ablation studies confirmed that the dropout layer is necessary and that fine-tuning its parameters can provide further benefits. The system further demonstrated robustness to multiple microphone failures. Finally, when the network size was reduced by two orders of magnitude, the resulting performance drop in SI-SDR was less than $2\,$dB.

These results demonstrate the generalization of the AmbiDrop framework across diverse array geometries and its suitability for deployment on resource-constrained edge devices such as wearables. Future research could include extending this framework to lightweight DNN architectures optimized for low-latency edge processing, and their real-time implementation.

\FloatBarrier
\bibliographystyle{IEEEtran}
\bibliography{references}

\end{document}